 \documentclass[preprint,12pt]{elsarticle}
\usepackage{amssymb,amsmath,latexsym}
\usepackage[dvips]{color}
\usepackage[headings]{fullpage}
\usepackage{epic,epsfig,graphicx}
\usepackage{enumerate}
\usepackage{multirow}
\usepackage{setspace}
\usepackage{keyval}
\usepackage{float}

\usepackage{amssymb}

 \newtheorem{thm}{Theorem}[section]

 \newtheorem{lemma}{Lemma}[section]
 \newtheorem{prop}[thm]{Proposition}
 \newtheorem{defn}{Definition}[section]

 \newtheorem{rem}{Remark}[section]
  \newtheorem{assumption}{Assumption}[section]

 \numberwithin{equation}{section}
\journal{***}

\begin{document}
\begin{frontmatter}
\title{The stochastic extinction and stability conditions for a class of malaria epidemic models}
\author{Divine Wanduku }
\address{Department of Mathematical Sciences,
Georgia Southern University, 65 Georgia Ave, Room 3042, Statesboro,
Georgia, 30460, U.S.A. E-mail:dwanduku@georgiasouthern.edu;wandukudivine@yahoo.com\footnote{Corresponding author. Tel: +14073009605.
} }
\begin{abstract}
 The stochastic extinction and stability in the mean of a family of SEIRS malaria models with a general nonlinear incidence rate is presented. The dynamics is driven by independent white noise processes from the disease transmission and natural death rates. The basic reproduction number $R^{*}_{0}$, the expected survival probability of the plasmodium $E(e^{-(\mu_{v}T_{1}+\mu T_{2})})$, and other threshold values are calculated. A sample Lyapunov exponential analysis for the system is utilized to obtain extinction results. Moreover, the rate of extinction of malaria is estimated, and innovative local Martingale and Lyapunov functional techniques are applied to establish the strong persistence, and asymptotic stability in the mean of the malaria-free steady population. 
  Moreover, for either $R^{*}_{0}<1$, or $E(e^{-(\mu_{v}T_{1}+\mu T_{2})})<\frac{1}{R^{*}_{0}}$, whenever  $R^{*}_{0}\geq 1$, respectively, extinction of malaria occurs. Furthermore, the robustness of these threshold conditions to the intensity of noise from the disease transmission rate is exhibited.  Numerical simulation results are presented.
\end{abstract}

\end{frontmatter}
\section{Introduction\label{ch1.sec0}}
In continuation with earlier discussions about malaria in Wanduku\cite{wanduku-biomath}, despite all technological advances to control the disease, malaria continues to exhibit an alarming high mortality rate. In fact, the latest WHO-\textit{World Malaria Report 2017} \cite{WHO-new} estimates a total of 216 million cases of malaria from 91 countries in 2016, which constitutes a 5 million increase in the total malaria cases from the malaria statistics obtained previously in 2015. Moreover, the total death count was 445000, and sub-Saharan Africa accounts for 90\% of the total estimated malaria cases. This rising prevalence trend in the malaria data continues to signal a need for more learning about the disease, improvement of the existing control strategies and equipment, and also a need for more advanced  resources etc. to fight  and eradicate, or ameliorate the burdens of the malaria.

 Malaria like other mosquito-borne diseases such as dengue fever, yellow fever, zika fever, lymphatic filariasis, and the different types of encephalitis etc. exhibits some  unique biological characteristics. For instance, the incubation of the disease requires two hosts - the mosquito vector and human hosts,  which may be either  directly involved in a full life cycle of the infectious agent consisting of two separate and independent segments of sub-life cycles which are completed separately inside the two hosts, or directly involved in two separate and independent half-life cycles of the infectious agent in the hosts. Therefore, there is a total latent time lapse of disease incubation which extends over the two segments of delay incubation times namely:- (1) the incubation period of the infectious agent ( or the half-life cycle) inside the vector, and (2) the incubation period of the infectious agent (or the other half-life cycle) inside the human being. See \cite{WHO,CDC}.

 Indeed,  the malaria plasmodium undergoes the first developmental half-life cycle called the \textit{sporogonic cycle} inside the female \textit{Anopheles} mosquito lasting approximately $10-18$ days, following a successful blood meal obtained from an infectious human being through a mosquito bite. Moreover, the mosquito  becomes infectious.  The parasite  completes the second developmental half-life cycle called the \textit{exo-erythrocytic cycle} lasting about 7-30 days inside the exposed human being\cite{WHO,CDC}, whenever the parasite is transferred to  human being in the process of the infectious mosquito foraging for another blood meal.

The exposure and successful recovery from a malaria parasite, for example, \textit{falciparum vivae} induces natural immunity against the disease which can protect against subsequent severe outbreaks of the disease. Moreover, the effectiveness and duration of the naturally acquired immunity against malaria is determined by several factors such as the  species and the frequency of  exposure to the parasites. Furthermore, it has been determined that other biological factors such as the genetics of the human being, for instance, sickle-cell anaemia, duffy negative blood types have bearings on the naturally acquired immunity against different species of malaria\cite{CDC,lars,denise}.

 Compartmental mathematical epidemic dynamic models  have been used to investigate the dynamics of several different types of  infectious diseases\cite{eric,sya,pang}.  In general, these models are classified as SIS, SIR, SIRS, SEIRS,  and  SEIR etc.\cite{qun,qunliu, nguyen,joaq, sena,wanduku-fundamental,Wanduku-2017, zhica} epidemic dynamic models depending on the compartments of the disease classes directly involved in the general disease dynamics.
  Some of these studies devote interest to SEIRS and SEIR models\cite{joaq,sena,cesar,sen,zhica,zheng}, which account for the compartment of individuals who are exposed to the disease, $E$, that is, infected but noninfectious individuals. This inclusion of the exposed class of individuals allows for more insights about the disease dynamics during the incubation stage of the disease.

 In addition, many of these epidemic dynamic models are improved in reality by including the time delays that occur in the disease dynamics. Generally, two distinct types of delays are studied namely:-disease latency and immunity delay. The disease latency represents the period of disease incubation, or period of infectiousness which nonetheless is studied as a delay in the disease dynamics. The immunity delay represents the period of effective naturally acquired immunity against the disease after successful recovery from infection. See \cite{Wanduku-2017,wanduku-delay,kyrychko,qun,zhica,cooke-driessche,shuj,Sampath,cooke, baretta-takeuchi1}.

 Stochastic epidemic dynamic models more realistically represent epidemic dynamic processes, because they include the randomness which naturally occurs during a disease outbreak, owing to the occurrence of random environmental fluctuations in the disease dynamics. The presence of noise in an epidemiological process may directly impact the state of the system -known as demographic white noise, or indirectly influence other driving parameters of the infectious system- known as environmental white noise. For example,  in \cite{Wanduku-2017,wanduku-fundamental,wanduku-delay},  environmental white noise is studied, where the noise represents the random fluctuations in the disease transmission rate. In \cite{qun}, the white noise process represents the variability in the natural death rate of the population. In \cite{Baretta-kolmanovskii},  the demographic white noise process represents the random fluctuations in the state of the system, where the noise deviates the state of the system from the equilibrium state, that is, the  white noise process is proportional to the difference between the state and equilibrium of the system. Some other authors such as \cite{allen,yicai} have suggested a mean-reverting process technique to include white noise processes.

  A stochastic white noise driven infectious system generally exhibits  more complex behavior in the disease dynamics, compared to their corresponding deterministic systems. For instance, the presence of noise  in the disease dynamics  may destabilize a disease-free steady state population, and drive the system into an endemic state. The occurrence of white noise with high intensity in the dynamics of a disease may continuously  decrease the population over time, leading to the extinction of the population. See for example \cite{qun,Wanduku-2017,wanduku-fundamental,wanduku-delay,zhuhu,yanli,qun}.

An important investigation in the study of infectious population dynamic systems influenced by white noise is the extinction of the disease, and the asymptotic stability of the disease-free population over sufficiently long time. Several papers in the literature\cite{aadil,yanli,yongli,yongli-2,yzhang,mao-2}  have addressed these topics. Investigations about the extinction of disease from the population seek to find conditions that favor the extinction of the disease-related classes such as the exposed and infectious classes in the population, and consequently lead to the survival of the susceptible and infection-free population classes over sufficiently long time. The techniques used to investigate extinction of the disease in stochastic systems include examining the sample paths of the system near a disease-free steady state, and computing the sample Lyapunov exponent of the trajectories of the system\cite{aadil,yanli,yongli,yongli-2,yzhang,mao-2}.

Cooke\cite{cooke} presented a deterministic epidemic dynamic model for  vector-borne diseases, where the bilinear incidence rate defined as $\beta S(t)I(t-T)$ represents the number of new infections occurring per unit time during the disease transmission process. It is assumed in the formulation of this incidence rate that the number of infectious vectors at time $t$ interacting and effectively transmitting infection to susceptible individuals, $S$, after $\beta$  number of effective contacts per unit time per infective is proportional to the infectious human population, $I$,  at earlier time $t-T$. Cook's method of effectively studying the dynamics of a vector-borne disease in a human population without directly including the vector population dynamics has been utilized by several other authors, for example \cite{baretta-takeuchi1,mcluskey,Wanduku-2017,qun,wanduku-biomath}.
%

Recently, Wanduku\cite{wanduku-biomath} presented and studied the following novel family of SEIRS epidemic dynamic models for malaria with three distributed delays:
%
\begin{equation}\label{ch1.sec0.eq3.intro.eq1}
\left\{
\begin{array}{lll}
dS(t)&=&\left[ B-\beta S(t)\int^{h_{1}}_{t_{0}}f_{T_{1}}(s) e^{-\mu s}G(I(t-s))ds - \mu S(t)+ \alpha \int_{t_{0}}^{\infty}f_{T_{3}}(r)I(t-r)e^{-\mu r}dr \right]dt,\\
dE(t)&=& \left[ \beta S(t)\int^{h_{1}}_{t_{0}}f_{T_{1}}(s) e^{-\mu s}G(I(t-s))ds - \mu E(t)\right.\\
&&\left.-\beta \int_{t_{0}}^{h_{2}}f_{T_{2}}(u)S(t-u)\int^{h_{1}}_{t_{0}}f_{T_{1}}(s) e^{-\mu s-\mu u}G(I(t-s-u))dsdu \right]dt,\\
dI(t)&=& \left[\beta \int_{t_{0}}^{h_{2}}f_{T_{2}}(u)S(t-u)\int^{h_{1}}_{t_{0}}f_{T_{1}}(s) e^{-\mu s-\mu u}G(I(t-s-u))dsdu- (\mu +d+ \alpha) I(t) \right]dt,\\
dR(t)&=&\left[ \alpha I(t) - \mu R(t)- \alpha \int_{t_{0}}^{\infty}f_{T_{3}}(r)I(t-r)e^{-\mu s}dr \right]dt,
\end{array}
\right.
\end{equation}
where the initial conditions are given in the following: let $h= h_{1}+ h_{2}$ and define
\begin{eqnarray}
&&\left(S(t),E(t), I(t), R(t)\right)
=\left(\varphi_{1}(t),\varphi_{2}(t), \varphi_{3}(t),\varphi_{4}(t)\right), t\in (-\infty,t_{0}],\nonumber\\
&&\varphi_{k}\in \mathcal{C}((-\infty,t_{0}],\mathbb{R}_{+}),\forall k=1,2,3,4,\quad \varphi_{k}(t_{0})>0,\forall k=1,2,3,4,\nonumber\\
 \label{ch1.sec0.eq3.intro.eq2}
\end{eqnarray}
where $\mathcal{C}((-\infty,t_{0}],\mathbb{R}_{+})$ is the space of continuous functions with  the supremum norm
\begin{equation}\label{ch1.sec0.eq3.intro.eq3}
||\varphi||_{\infty}=\sup_{ t\leq t_{0}}{|\varphi(t)|}.
\end{equation}
The disease spreads in the human population of total size $ N(t)=S(t)+ E(t)+ I(t)+ R(t)$, where $S(t)$, $E(t)$, $I(t)$ and $R(t)$ represent the susceptible, exposed, infectious and naturally acquired immunity classes at time $t$, respectively. The positive constants $B$, and $\mu$ represent the constant birth and natural death rates, respectively. Furthermore, the disease related deathrate is denoted $d$. The rate $\beta$ is the average effective contact rate per infected mosquito per unit time. The recovery rate from malaria with acquired immunity is $\alpha$. Also, the incubation delays inside the mosquito and human hosts are denoted $T_{1}$ and $T_{2}$, respectively, and the period of effective naturally acquired immunity is denoted $T_{3}$. Moreover, the delays are random variables with arbitrary densities denoted $f_{T_{1}}$, $f_{T_{2}}$ and $f_{T_{3}}$, and their supports given as $T_{1}\in[t_{0},h_{1}]$, $T_{2}\in[t_{0},h_{1}]$ and $T_{3}\in[t_{0}, +\infty)$.
The nonlinear incidence function $G$ which signifies the response to disease transmission by the susceptible class as malaria increases in the population, satisfies the following assumptions  \begin{assumption}\label{ch1.sec0.assum1}
\begin{enumerate}
  \item [$A1$]$G(0)=0$; $A2$: $G(I)$ is strictly monotonic on $[0,\infty)$; $A3$: $G''(I)<0$;
   $A4$. $\lim_{I\rightarrow \infty}G(I)=C, 0\leq C<\infty$; 
  and  $A5$: $G(I)\leq I, \forall I>0$.
\end{enumerate}
\end{assumption}
More details about the derivation of the model in (\ref{ch1.sec0.eq3}) is given in Wanduku\cite{wanduku-biomath}.  The study Wanduku\cite{wanduku-biomath} provides a suitable platform to investigate the dynamics of mosquito-borne diseases of humans with similar general structure as malaria, for instance, dengue fever, yellow fever, zika fever, lymphatic filariasis, and the different types of encephalitis etc. which are all transmitted by the mosquito. Moreover, the study \cite{wanduku-biomath} also provides a  technique to incorporate the multiple general forms of delays in the vector-borne disease dynamics namely-disease latency and immunity delay, and investigate the dynamics of vector-borne diseases without directly including the dynamics of the vector population.

In the analysis of the deterministic malaria model (\ref{ch1.sec0.eq3.intro.eq1}) with initial conditions in (\ref{ch1.sec0.eq3.intro.eq2})-(\ref{ch1.sec0.eq3.intro.eq3})  in Wanduku\cite{wanduku-biomath}, the threshold values for disease eradication such as the basic reproduction number for the disease when the system is in steady state are obtained in both cases where the delays in the system $T_{1}, T_{2}$ and $T_{3}$ are constant, and  also arbitrarily distributed.

For $S^{*}_{0}=\frac{B}{\mu}$, when the delays in the system are all constant, the basic reproduction number of the disease is given by
\begin{equation} \label{ch1.sec2.lemma2a.corrolary1.eq4}
\hat{R}^{*}_{0}=\frac{\beta S^{*}_{0} }{(\mu+d+\alpha)}.
\end{equation}
Furthermore, the threshold condition $\hat{R}^{*}_{0}<1$ is required for the disease-free equilibrium $E_{0}=(S^{*}_{0},0,0)$ to be asymptotically stable, and for the disease  to be eradicated from the steady state human population.

On the other hand, when the delays in the system  $T_{i}, i=1,2,3$  are random, and arbitrarily distributed, the basic reproduction number is given by
\begin{equation}\label{ch1.sec2.theorem1.corollary1.eq3}
R_{0}=\frac{\beta S^{*}_{0} \hat{K}_{0}}{(\mu+d+\alpha)}+\frac{\alpha}{(\mu+d+\alpha)},
\end{equation}
where, $\hat{K}_{0}>0$ is a constant that depends only on $S^{*}_{0}$  (in fact, $\hat{K}_{0}=4+ S^{*}_{0} $). In addition, malaria is eradicated from the system in the steady state, whenever $R_{0}\leq 1$,

 The results in [Theorem~5.1, Wanduku\cite{wanduku-biomath}] also show that when $R_{0}>1$, and the expected survival probability rate $E(e^{-\mu(T_{1}+ T_{2})})$ of the parasites over their complete life cycle is significant,  the deterministic system (\ref{ch1.sec0.eq3}) establishes a unique endemic equilibrium state denoted by $E_{1}=(S^{*}_{1}, E^{*}_{1}, I^{*}_{1})$.

The current paper extends the previous study Wanduku\cite{wanduku-biomath}, by incorporating the independent white noise perturbations of the effective disease transmission rate $\beta$, and the natural deathrates of the susceptible, exposed, infectious and removal populations.
%
  The primary focus of this study is to investigate the extinction of malaria in a class of stochastic models for vector-borne diseases in a very noisy environment comprising of variability from the disease transmission and natural death rates.

It is important to note that this study is part of the broader project investigating vector-borne diseases in the human population. As part of this project, a deterministic study of malaria has already appeared in  Wanduku\cite{wanduku-biomath}. Some specialized stochastic extensions of this project addressing the impacts of noise on the persistence of malaria in the endemic equilibrium population will appear in Wanduku\cite{wanduku-comparative}. Moreover, the stochastic permanence of malaria and existence of stationary distribution will appear in Wanduku\cite{wanduku-permanence}.

   This work is presented as follows:- in Section~\ref{ch1.sec0}, the epidemic dynamic model is derived. In Section~\ref{ch1.sec1}, the model validation results are presented. In Section~\ref{ch1.sec2a}, the extinction conditions for the disease are presented for the case where the noise stems jointly from the disease transmission and natural death rates.   In Section~\ref{ch1.sec2b}, the extinction conditions for the disease are presented for the case where the noise stems only from the disease transmission rate. Moreover, the asymptotic stability in the mean of the disease-free equilibrium is also presented.  Finally, in Section~\ref{ch1.sec4},  numerical simulation results are given.
\section{Derivation of the stochastic Model}\label{ch1.sec0}
 The class of malaria models in Wanduku\cite{wanduku-biomath} is adapted  and extended into a generalized class of stochastic SEIRS delayed epidemic dynamic models for vector-borne diseases in the following. The delays represent the incubation period of the infectious agents (plasmodium or dengue fever virus etc.) in the vector $T_{1}$, and in the human host $T_{2}$. The third delay represents the naturally acquired immunity period of the disease $T_{3}$, where the delays are random variables with density functions $f_{T_{1}}, t_{0}\leq T_{1}\leq h_{1}, h_{1}>0$, and $f_{T_{2}}, t_{0}\leq T_{2}\leq h_{2}, h_{2}>0$ and $f_{T_{3}}, t_{0}\leq T_{3}<\infty$. All other assumptions for $T_{1}, T_{2}$ and $T_{3}$ are similar to the study \cite{wanduku-biomath}.
  %

By employing similar reasoning in \cite{cooke,qun,capasso,huo}, the expected incidence rate of the disease or force of infection of the disease at time $t$ due to the disease transmission process between the infectious vectors and susceptible humans, $S(t)$, is given by the expression $\beta \int^{h_{1}}_{t_{0}}f_{T_{1}}(s) e^{-\mu_{v} s}S(t)G(I(t-s))ds$, where $\mu_{v}$ is the natural death rate of the vectors in the population. Assuming exponential lifetime for the random incubation period $T_{1}$, the probability rate, $0<e^{-\mu_{v} s}\leq 1, s\in [t_{0}, h_{1}], h_{1}>0$,  represents the survival probability rate of  exposed vectors over the incubation period, $T_{1}$, of the infectious agent inside the vectors with the length of the period given as $T_{1}=s, \forall s \in [t_{0}, h_{1}]$, where the vectors acquired infection at the earlier time $t-s$  from an infectious human via for instance, biting and collecting an infected blood meal, and  become infectious at time $t$.  Furthermore, it is assumed that the survival of the vectors over the incubation period of length  $s\in [t_{0}, h_{1}]$ is independent of the age of the vectors. In addition, $I(t-s)$, is the infectious human population at earlier time $t-s$, $G$ is a nonlinear incidence function for the disease dynamics,  and $\beta$ is the average number of effective contacts per infectious individual per unit time. Indeed, the force of infection,  $\beta \int^{h_{1}}_{t_{0}}f_{T_{1}}(s) e^{-\mu_{v} s}S(t)G(I(t-s))ds$  signifies the expected rate of new infections at time $t$ between the infectious vectors and the susceptible human population $S(t)$ at time $t$, where the infectious agent is transmitted per infectious vector per unit time at the rate $\beta$. Furthermore, it is assumed that the number of infectious vectors at time $t$ is proportional to the infectious human population at earlier time $t-s$ (see \cite{cooke}). Moreover, it is further assumed that the interaction between the infectious vectors and  susceptible humans exhibits nonlinear behavior, for instance, psychological and overcrowding effects,  which is characterized by the nonlinear incidence  function $G$. Therefore, the force of infection given by
 \begin{equation}\label{ch1.sec0.eqn0}
   \beta \int^{h_{1}}_{t_{0}}f_{T_{1}}(s) e^{-\mu_{v} s}S(t)G(I(t-s))ds,
 \end{equation}
  represents the expected rate at which infected individuals leave the susceptible state and become exposed at time $t$.

It is assumed that the natural death of human beings in the population is $\mu$. From a biological point of view, the average lifespan of vectors $\frac{1}{\mu_{v}}$, is much less than the average lifespan of a human being in the absence of disease $\frac{1}{\mu}$. It follows very easily that assuming exponential lifetime for the random variables  $T_{1}=s\in [t_{0},h_{1}]$, and $T_{2}=s\in [t_{0},h_{2}]$, the survival probabilities satisfy
\begin{equation}\label{ch1.sec0.eqn0.eq1}
e^{-\mu_{v}T_{1}}<<e^{-\mu T_{1}}\quad and\quad e^{-\mu_{v}T_{1}-\mu T_{2}}<<e^{-\mu(T_{1} +T_{2})}.
\end{equation}
That is, (\ref{ch1.sec0.eqn0.eq1}) signifies that the survival chance of the mosquitoes, and consequently the parasites or virus inside the mosquitoes over the complete life cycle of the parasites lasting for $T_{1}+T_{2}$ time units, is less than the survival chance of human beings over the same period of time. Furthermore, recall [Theorem~5.1, \cite{wanduku-biomath}] asserts that it is necessary for the expected survival rate $E(e^{-\mu_{v}T_{1}-\mu T_{2}})$ to be significant for the disease to establish a steady endemic population.

 The susceptible individuals who have acquired infection from infectious vectors but are non infectious form the exposed class $E$. The population of exposed individuals at time $t$ is denoted $E(t)$. After the incubation period, $T_{2}=u\in [t_{0}, h_{2}]$, of the infectious agent in the exposed human host, the individual becomes infectious, $I(t)$, at time $t$. Applying similar reasoning in  \cite{cooke-driessche},
 the exposed population, $E(t)$, at time $t$ can be written as follows
  \begin{equation}\label{ch1.sec0.eqn1a}
    E(t)=E(t_{0})e^{-\mu (t-t_{0})}p_{1}(t-t_{0})+\int^{t}_{t_{0}}\beta S(\xi)e^{-\mu_{v} T_{1}}G(I(\xi-T_{1}))e^{-\mu(t-\xi)}p_{1}(t-\xi)d \xi,
   \end{equation}
   where
   \begin{equation}\label{ch1.seco.eqn1b}
     p_{1}(t)=\left\{\begin{array}{l}0,t\geq T_{2},\\
 1, t< T_{2} \end{array}\right.
   \end{equation}
   represents the probability that an individual remains exposed over the time interval $[0, t]$.
   It is easy to see from (\ref{ch1.sec0.eqn1a}) that under the assumption that the disease has been in the population for at least a time $t>\max_{t_{0}\leq T_{1}\leq h_{1}, t_{0}\leq T_{2}\leq h_{2}} {( T_{1}+ T_{2})}$, in fact, $t>h_{1}+h_{2}$, so that all initial perturbations have died out, the expected number of exposed individuals at time $t$ is given  by
\begin{equation}\label{ch1.sec0.eqn1}
E(t)=\int_{t_{0}}^{h_{2}}f_{T_{2}}(u)\int_{t-u}^{t}\beta \int^{h_{1}}_{t_{0}} f_{T_{1}}(s) e^{-\mu_{v} s}S(v)G(I(v-s))e^{-\mu(t-u)}dsdvdu.
\end{equation}
    Similarly, for the removal population, $R(t)$, at time $t$, individuals recover from the infectious state $I(t)$  at the per capita rate $\alpha$  and acquire natural immunity.  The natural immunity wanes after the varying immunity period $T_{3}=r\in [ t_{0},\infty]$, and removed individuals become susceptible again to the disease. Therefore, at time $t$, individuals leave the infectious state at the rate $\alpha I(t)$  and become part of the removal population $R(t)$. Thus, at time $t$ the removed population is given by the following equation
  \begin{equation}\label{ch1.sec0.eqn2a}
    R(t)=R(t_{0})e^{-\mu (t-t_{0})}p_{2}(t-t_{0})+\int^{t}_{t_{0}}\alpha I(\xi)e^{-\mu(t-\xi)}p_{2}(t-\xi)d \xi,
  \end{equation}
    where
    \begin{equation}\label{ch1.sec0.eqn2b}
      p_{2}(t)=\left\{\begin{array}{l}0,t\geq T_{3},\\
 1, t< T_{3} \end{array}\right.
    \end{equation}
 represents the probability that an individual remains naturally immune to the disease over the time interval $[0, t]$.
 But it follows from  (\ref{ch1.sec0.eqn2a}) that under the assumption that the disease has been in the population for at least a time $t> \max_{t_{0}\leq T_{1}\leq h_{1}, t_{0}\leq T_{2}\leq h_{2}, T_{3}\geq t_{0}}{(T_{1}+ T_{2}, T_{3})}\geq \max_{ T_{3}\geq t_{0}}{(T_{3})}$, in fact, the disease has been in the population for sufficiently large  amount of time so that all initial perturbations have died out,  then the expected number of removal individuals at time $t$ can be written as
  \begin{equation}\label{ch1.sec0.eqn2}
R(t)=\int_{t_{0}}^{\infty}f_{T_{3}}(r)\int_{t-r}^{t}\alpha I(v)e^{-\mu (t-v)}dvdr.
\end{equation}
There is also constant birth rate $B$ of susceptible individuals in the population. Furthermore, individuals die additionally due to disease related causes at the rate $d$. All otherr assumptions for the malaria model (\ref{ch1.sec0.eq3.intro.eq1}) remain the same.
It follows from (\ref{ch1.sec0.eqn0}), (\ref{ch1.sec0.eqn1}), (\ref{ch1.sec0.eqn2}) that the SEIRS malaria model (\ref{ch1.sec0.eq3.intro.eq1}) can be written in a more detailed form as follows
\begin{eqnarray}
dS(t)&=&\left[ B-\beta S(t)\int^{h_{1}}_{t_{0}}f_{T_{1}}(s) e^{-\mu_{v} s}G(I(t-s))ds - \mu S(t)+ \alpha \int_{t_{0}}^{\infty}f_{T_{3}}(r)I(t-r)e^{-\mu r}dr \right]dt,\nonumber\\
&&\label{ch1.sec0.eq3}\\
dE(t)&=& \left[ \beta S(t)\int^{h_{1}}_{t_{0}}f_{T_{1}}(s) e^{-\mu_{v} s}G(I(t-s))ds - \mu E(t)\right.\nonumber\\
&&\left.-\beta \int_{t_{0}}^{h_{2}}f_{T_{2}}(u)S(t-u)\int^{h_{1}}_{t_{0}}f_{T_{1}}(s) e^{-\mu_{v} s-\mu u}G(I(t-s-u))dsdu \right]dt,\label{ch1.sec0.eq4}\\
&&\nonumber\\
dI(t)&=& \left[\beta \int_{t_{0}}^{h_{2}}f_{T_{2}}(u)S(t-u)\int^{h_{1}}_{t_{0}}f_{T_{1}}(s) e^{-\mu_{v} s-\mu u}G(I(t-s-u))dsdu- (\mu +d+ \alpha) I(t) \right]dt,\nonumber\\
&&\label{ch1.sec0.eq5}\\
dR(t)&=&\left[ \alpha I(t) - \mu R(t)- \alpha \int_{t_{0}}^{\infty}f_{T_{3}}(r)I(t-r)e^{-\mu s}dr \right]dt.\label{ch1.sec0.eq6}
\end{eqnarray}
Furthermore, the incidence function $G$ satisfies the assumptions in Assumption~\ref{ch1.sec0.assum1}.

 It is assumed in the current study that the  effects of random environmental fluctuations lead to variability in the disease transmission and natural death rates.
 For $t\geq t_{0}$, let $(\Omega, \mathfrak{F}, P)$ be a complete probability space, and $\mathfrak{F}_{t}$ be a filtration (that is, sub $\sigma$- algebra $\mathfrak{F}_{t}$ that satisfies the following: given $t_{1}\leq t_{2} \Rightarrow \mathfrak{F}_{t_{1}}\subset \mathfrak{F}_{t_{2}}; E\in \mathfrak{F}_{t}$ and $P(E)=0 \Rightarrow E\in \mathfrak{F}_{0} $ ).
 The variability in the disease transmission and natural death rates are represented by independent white noise processes, and the rates are expressed as follows:
 \begin{equation}\label{ch1.sec0.eq7}
 \mu  \rightarrow \mu  + \sigma_{i}\xi_{i}(t),\quad \xi_{i}(t) dt= dw_{i}(t),i=S,E,I,R, \quad  \beta \rightarrow \beta + \sigma_{\beta}\xi_{\beta}(t),\quad \xi_{\beta}(t)dt=dw_{\beta}(t),
 \end{equation}
 where $\xi_{i}(t)$ and $w_{i}(t)$ represent the  standard white noise and normalized wiener processes  for the  $i^{th}$ state at time $t$, with the following properties: $w(0)=0, E(w(t))=0, var(w(t))=t$.  Furthermore,  $\sigma_{i},i=S,E,I,R $, represents the intensity value of the white noise process due to the natural death rate in the $i^{th}$ state, and $\sigma_{\beta}$ is the intensity value of the white noise process due to the disease transmission rate.

  The ideas behind the formulation of the expressions in (\ref{ch1.sec0.eq7}) are given in the following. The constant parameters $\mu$ and $\beta$ represent the natural death and disease transmission rates per unit time, respectively. In reality, random environmental fluctuations impact these rates turning them into random variables $\tilde{\mu}$ and $\tilde{\beta}$. Thus, the natural death and disease transmission rates over an infinitesimally small interval of time $[t, t+dt]$  with length $dt$ is given by the expressions  $\tilde{\mu}(t)=\tilde{\mu}dt$ and $\tilde{\beta}(t)=\tilde{\beta}dt$, respectively. It is assumed that there are independent and identical random impacts acting upon these rates at times $t_{j+1}$ over $n$ subintervals $[t_{j}, t_{j+1}]$ of length $\triangle t=\frac{dt}{n}$, where $t_{j}=t_{0}+j\triangle t, j=0,1,\cdots,n$, and  $t_{0}=t$. Furthermore,  it is assumed that $\tilde{\mu}(t_{0})=\tilde{\mu}(t)=\mu dt$ is  constant or deterministic, and $\tilde{\beta}(t_{0})=\tilde{\beta}(t)=\beta dt$ is also a constant. It follows that by letting the independent identically distributed random variables $Z_{i},i=1,\cdots,n $ represent the random effects acting on the natural death rate, then it follows further that the rate at time $t_{n}=t+dt$, that is,
   \begin{equation}\label{ch1.sec0.eq7.eq1}
    \tilde{\mu}(t+dt)=\tilde{\mu}(t)+\sum_{j=1}^{n}Z_{j},
  \end{equation}
  where $E(Z_{j})=0$,and $Var(Z_{j})=\sigma^{2}_{i}\triangle t, i\in \{S, E, I, R\}$.
    Note that $\tilde{\beta}(t+dt)$ can similarly be expressed as (\ref{ch1.sec0.eq7.eq1}). And for sufficient large value of $n$, the summation in (\ref{ch1.sec0.eq7.eq1}) converges in distribution by the central limit theorem to a random variable which is identically distributed as the wiener process $\sigma_{i}(w_{i}(t+dt)-w_{i}(t))=\sigma_{i}dw_{i}(t)$, with mean $0$ and variance $\sigma^{2}_{i}dt, i\in \{S, E, I, R\}$. It follows easily from (\ref{ch1.sec0.eq7.eq1}) that
  \begin{equation}\label{ch1.sec0.eq7.eq2}
   \tilde{\mu}dt =\mu dt+ \sigma_{i}dw_{i}(t), i\in \{S, E, I, R\}.
  \end{equation}
  Similarly, it can be easily seen that
  \begin{equation}\label{ch1.sec0.eq7.eq2}
   \tilde{\beta}dt =\beta dt+ \sigma_{\beta}dw_{\beta}(t).
  \end{equation}
    Note that the intensities $\sigma^{2}_{i},i=S,E,I,R, \beta $ of the independent white noise processes in the expressions   $\tilde{\mu}(t)=\mu dt  + \sigma_{i}\xi_{i}(t)$ and $\tilde{\beta} (t)=\beta dt + \sigma_{\beta}\xi_{\beta}(t)$ that represent the  natural death rate, $\tilde{\mu}(t)$, and disease transmission rate, $\tilde{\beta} (t)$,  at time $t$,  measure the average deviation of the random variable disease transmission rate, $\tilde{\beta}$,  and natural death rate, $\tilde{\mu}$, about their constant mean values $\beta$ and $\mu$, respectively, over the infinitesimally small time interval $[t, t+dt]$. These measures reflect the force of the random fluctuations that occur during the disease outbreak at anytime, and which lead to oscillations in the natural death and disease transmission rates overtime, and consequently lead to oscillations of the susceptible, exposed, infectious and removal states of the system over time during the disease outbreak. Thus, in this study the words "strength" and "intensity" of the white noise are used synonymously. Also, the constructions "strong noise" and "weak noise" are used to refer to white noise with high and low intensities, respectively.

    Under the assumptions in the formulation of the natural death rate per unit time $\tilde{\mu}$ as a brownian motion process above, it can also be seen  easily that under further assumption that the number of natural deaths $N(t)$ over an interval $[t_{0}, t_{0}+t]$ of length $t$ follows a poisson process $\{N(t),t\geq t_{0}\}$ with intensity of the process $E(\tilde{\mu})=\mu$, and mean $E(N(t))=E(\tilde{\mu}t)=\mu t$, then the lifetime is exponentially distributed with mean $\frac{1}{\mu}$ and survival function
    \begin{equation}\label{ch1.sec0.eq7.eq3}
      \mathfrak{S}(t)=e^{-\mu t},t>0.
    \end{equation}
 Substituting (\ref{ch1.sec0.eq7})-(\ref{ch1.sec0.eq7.eq3}) into the deterministic system (\ref{ch1.sec0.eq3})-(\ref{ch1.sec0.eq6}) leads to the following generalized system of Ito-Doob stochastic differential equations describing the dynamics of  vector-borne diseases in the human population.
 \begin{eqnarray}
dS(t)&=&\left[ B-\beta S(t)\int^{h_{1}}_{t_{0}}f_{T_{1}}(s) e^{-\mu_{v} s}G(I(t-s))ds - \mu S(t)+ \alpha \int_{t_{0}}^{\infty}f_{T_{3}}(r)I(t-r)e^{-\mu r}dr \right]dt\nonumber\\
&&-\sigma_{S}S(t)dw_{S}(t)-\sigma_{\beta} S(t)\int^{h_{1}}_{t_{0}}f_{T_{1}}(s) e^{-\mu_{v} s}G(I(t-s))dsdw_{\beta}(t) \label{ch1.sec0.eq8}\\
dE(t)&=& \left[ \beta S(t)\int^{h_{1}}_{t_{0}}f_{T_{1}}(s) e^{-\mu_{v} s}G(I(t-s))ds - \mu E(t)\right.\nonumber\\
&&\left.-\beta \int_{t_{0}}^{h_{2}}f_{T_{2}}(u)S(t-u)\int^{h_{1}}_{t_{0}}f_{T_{1}}(s) e^{-\mu_{v} s-\mu u}G(I(t-s-u))dsdu \right]dt\nonumber\\
&&-\sigma_{E}E(t)dw_{E}(t)+\sigma_{\beta} S(t)\int^{h_{1}}_{t_{0}}f_{T_{1}}(s) e^{-\mu_{v} s}G(I(t-s))dsdw_{\beta}(t)\nonumber\\
&&-\sigma_{\beta} \int_{t_{0}}^{h_{2}}f_{T_{2}}(u)S(t-u)\int^{h_{1}}_{t_{0}}f_{T_{1}}(s) e^{-\mu_{v} s-\mu u}G(I(t-s-u))dsdudw_{\beta}(t)\label{ch1.sec0.eq9}\\
dI(t)&=& \left[\beta \int_{t_{0}}^{h_{2}}f_{T_{2}}(u)S(t-u)\int^{h_{1}}_{t_{0}}f_{T_{1}}(s) e^{-\mu_{v} s-\mu u}G(I(t-s-u))dsdu- (\mu +d+ \alpha) I(t) \right]dt\nonumber\\
&&-\sigma_{I}I(t)dw_{I}(t)+\sigma_{\beta} \int_{t_{0}}^{h_{2}}f_{T_{2}}(u)S(t-u)\int^{h_{1}}_{t_{0}}f_{T_{1}}(s) e^{-\mu_{v} s-\mu u}G(I(t-s-u))dsdudw_{\beta}(t)\nonumber\\
&&\label{ch1.sec0.eq10}\\
dR(t)&=&\left[ \alpha I(t) - \mu R(t)- \alpha \int_{t_{0}}^{\infty}f_{T_{3}}(r)I(t-r)e^{-\mu s}dr \right]dt-\sigma_{R}R(t)dw_{R}(t),\label{ch1.sec0.eq11}
\end{eqnarray}
where the initial conditions are given in the following:- where ever necessary, we  let $h= h_{1}+ h_{2}$ and define
\begin{eqnarray}
&&\left(S(t),E(t), I(t), R(t)\right)
=\left(\varphi_{1}(t),\varphi_{2}(t), \varphi_{3}(t),\varphi_{4}(t)\right), t\in (-\infty,t_{0}],\nonumber\\
&&\varphi_{k}\in \mathcal{C}((-\infty,t_{0}],\mathbb{R}_{+}),\forall k=1,2,3,4, \nonumber\\
&&\varphi_{k}(t_{0})>0,\forall k=1,2,3,4,\nonumber\\
 \label{ch1.sec0.eq12}
\end{eqnarray}
where $\mathcal{C}((-\infty,t_{0}],\mathbb{R}_{+})$ is the space of continuous functions with  the supremum norm
\begin{equation}\label{ch1.sec0.eq13}
||\varphi||_{\infty}=\sup_{ t\leq t_{0}}{|\varphi(t)|}.
\end{equation}
Furthermore, the random continuous functions $\varphi_{k},k=1,2,3,4$ are
$\mathfrak{F}_{0}-measurable$, or  independent of $w(t)$
for all $t\geq t_{0}$.

It can be observed that (\ref{ch1.sec0.eq9}) and (\ref{ch1.sec0.eq11}) decouple from the other equations for $S$ and $I$ in the system (\ref{ch1.sec0.eq8})-(\ref{ch1.sec0.eq11}). It is customary to show the results for this kind of decoupled system using the simplified system containing only the non-decoupled system equations for $S$ and $I$, and then infer the results to the states $E$ and $R$, since these states depend exclusively on $S$ and $I$.  Nevertheless, for convenience, the existence  results of the system (\ref{ch1.sec0.eq8})-(\ref{ch1.sec0.eq11}) will be shown for the vector $X(t)=(S(t), E(t), I(t))$, and the extinction results presented for the decoupled system $(S(t),I(t))$. The following notations will be used throughout this study:
\begin{equation}\label{ch1.sec0.eq13b}
\left\{
  \begin{array}{lll}
    Y(t)&=&(S(t), E(t), I(t), R(t)) \\
   X(t)&=&(S(t), E(t), I(t)) \\
   N(t)&=&S(t)+ E(t)+ I(t)+ R(t).
  \end{array}
  \right.
\end{equation}
\section{Model Validation Results\label{ch1.sec1}}
In this section, the existence and uniqueness results for the solutions of the stochastic system  (\ref{ch1.sec0.eq8})-(\ref{ch1.sec0.eq11}) are presented.  The standard method  utilized in the earlier studies\cite{Wanduku-2017,wanduku-delay,divine5} is applied to establish the results.
  It should be noted that the existence and qualitative behavior of the positive solutions of the system (\ref{ch1.sec0.eq8})-(\ref{ch1.sec0.eq11}) depend on the sources (natural death or disease transmission rates) of variability in the system. As it is shown below, certain sources of variability lead to very complex uncontrolled behavior of the solutions of the system.
  The following Lemma describes the behavior of the positive local solutions for the system (\ref{ch1.sec0.eq8})-(\ref{ch1.sec0.eq11}). This result will be useful in   establishing the existence and uniqueness results for the global solutions of the stochastic system (\ref{ch1.sec0.eq8})-(\ref{ch1.sec0.eq11}).
\begin{lemma}\label{ch1.sec1.lemma1}
Suppose for some $\tau_{e}>t_{0}\geq 0$ the system (\ref{ch1.sec0.eq8})-(\ref{ch1.sec0.eq11}) with initial condition in (\ref{ch1.sec0.eq12}) has a unique positive solution $Y(t)\in \mathbb{R}^{4}_{+}$, for all $t\in (-\infty, \tau_{e}]$, then  if $N(t_{0})\leq \frac{B}{\mu}$, it follows that
\item[(a.)] if the intensities of the independent white noise processes in the system satisfy  $\sigma_{i}=0$, $i\in \{S, E, I\}$ and $\sigma_{\beta}\geq 0$, then $N(t)\leq \frac{B}{\mu}$, and in addition, the set denoted by
\begin{equation}\label{ch1.sec1.lemma1.eq1}
  D(\tau_{e})=\left\{Y(t)\in \mathbb{R}^{4}_{+}: N(t)=S(t)+ E(t)+ I(t)+ R(t)\leq \frac{B}{\mu}, \forall t\in (-\infty, \tau_{e}] \right\}=\bar{B}^{(-\infty, \tau_{e}]}_{\mathbb{R}^{4}_{+},}\left(0,\frac{B}{\mu}\right),
\end{equation}
is locally self-invariant with respect to the system (\ref{ch1.sec0.eq8})-(\ref{ch1.sec0.eq11}), where $\bar{B}^{(-\infty, \tau_{e}]}_{\mathbb{R}^{4}_{+},}\left(0,\frac{B}{\mu}\right)$ is the closed ball in $\mathbb{R}^{4}_{+}$ centered at the origin with radius $\frac{B}{\mu}$ containing the local positive solutions defined over $(-\infty, \tau_{e}]$.
\item[(b.)] If the intensities of the independent white noise processes in the system satisfy  $\sigma_{i}>0$, $i\in \{S, E, I\}$ and $\sigma_{\beta}\geq 0$, then $N(t)\geq 0$, for all $t\in (-\infty, \tau_{e}]$.
\end{lemma}
Proof:\\
 It follows directly from (\ref{ch1.sec0.eq8})-(\ref{ch1.sec0.eq11}) that when $\sigma_{i}=0$, $i\in \{S, E, I\}$ and $\sigma_{\beta}\geq 0$, then
\begin{equation}\label{ch1.sec1.lemma1.eq2}
dN(t)=[B-\mu N(t)-dI(t)]dt
\end{equation}
The result in (a.) follows easily by observing that for $Y(t)\in \mathbb{R}^{4}_{+}$, the equation (\ref{ch1.sec1.lemma1.eq2}) leads to  $N(t)\leq \frac{B}{\mu}-\frac{B}{\mu}e^{-\mu(t-t_{0})}+N(t_{0})e^{-\mu(t-t_{0})}$. And under the assumption that $N(t_{0})\leq \frac{B}{\mu}$, the result follows immediately. The result in (b.) follows directly from Theorem~\ref{ch1.sec1.thm1}.
 The following theorem presents the existence and uniqueness results for the global solutions  of the stochastic system (\ref{ch1.sec0.eq8})-(\ref{ch1.sec0.eq11}). 
\begin{thm}\label{ch1.sec1.thm1}
  Given the initial conditions (\ref{ch1.sec0.eq12}) and (\ref{ch1.sec0.eq13}), there exists a unique solution process $X(t,w)=(S(t,w),E(t,w), I(t,w))^{T}$ satisfying (\ref{ch1.sec0.eq8})-(\ref{ch1.sec0.eq11}), for all $t\geq t_{0}$. Moreover,
  \item[(a.)] the solution process is positive for all $t\geq t_{0}$ a.s. and lies in $D(\infty)$, whenever  the intensities of the independent white noise processes in the system satisfy  $\sigma_{i}=0$, $i\in \{S, E, I\}$ and $\sigma_{\beta}\geq 0$.
        That is, $S(t,w)>0,E(t,w)>0,  I(t,w)>0, \forall t\geq t_{0}$ a.s. and $X(t,w)\in D(\infty)=\bar{B}^{(-\infty, \infty)}_{\mathbb{R}^{4}_{+},}\left(0,\frac{B}{\mu}\right)$, where $D(\infty)$ is defined in Lemma~\ref{ch1.sec1.lemma1}, (\ref{ch1.sec1.lemma1.eq1}).
        \item[(b.)] Also, the solution process is positive for all $t\geq t_{0}$ a.s. and lies in $\mathbb{R}^{4}_{+}$, whenever  the intensities of the independent white noise processes in the system satisfy  $\sigma_{i}>0$, $i\in \{S, E, I\}$ and $\sigma_{\beta}\geq 0$.
        That is, $S(t,w)>0,E(t,w)>0,  I(t,w)>0, \forall t\geq t_{0}$ a.s. and $X(t,w)\in \mathbb{R}^{4}_{+}$.
\end{thm}
Proof:\\
It is easy to see that the coefficients of (\ref{ch1.sec0.eq8})-(\ref{ch1.sec0.eq11}) satisfy the local Lipschitz condition for the given initial data (\ref{ch1.sec0.eq12}). Therefore there exist a unique maximal local solution $X(t,w)=(S(t,w), E(t,w), I(t,w))$ on $t\in (-\infty,\tau_{e}(w)]$, where $\tau_{e}(w)$ is the first hitting time or the explosion time\cite{mao}. The following shows that $X(t,w)\in D(\tau_{e})$ almost surely, whenever $\sigma_{i}=0$, $i\in \{S, E, I\}$ and $\sigma_{\beta}\geq 0$,  where $D(\tau_{e}(w))$ is defined in Lemma~\ref{ch1.sec1.lemma1} (\ref{ch1.sec1.lemma1.eq1}), and also that $X(t,w)\in \mathbb{R}^{4}_{+}$, whenever  $\sigma_{i}>0$, $i\in \{S, E, I\}$ and $\sigma_{\beta}\geq 0$.
Define the following stopping time
\begin{equation}\label{ch1.sec1.thm1.eq1}
\left\{
\begin{array}{lll}
\tau_{+}&=&\sup\{t\in (t_{0},\tau_{e}(w)): S|_{[t_{0},t]}>0,\quad E|_{[t_{0},t]}>0,\quad and\quad I|_{[t_{0},t]}>0 \},\\
\tau_{+}(t)&=&\min(t,\tau_{+}),\quad for\quad t\geq t_{0}.\\
\end{array}
\right.
\end{equation}
and lets show that $\tau_{+}(t)=\tau_{e}(w)$ a.s. Suppose on the contrary that $P(\tau_{+}(t)<\tau_{e}(w))>0$. Let $w\in \{\tau_{+}(t)<\tau_{e}(w)\}$, and $t\in [t_{0},\tau_{+}(t))$. Define
\begin{equation}\label{ch1.sec1.thm1.eq2}
\left\{
\begin{array}{ll}
V(X(t))=V_{1}(X(t))+V_{2}(X(t))+V_{3}(X(t)),\\
V_{1}(X(t))=\ln(S(t)),\quad V_{2}(X(t))=\ln(E(t)),\quad V_{3}(X(t))=\ln(I(t)),\forall t\leq\tau_{+}(t).
\end{array}
\right.
\end{equation}
It follows from (\ref{ch1.sec1.thm1.eq2}) that
\begin{equation}\label{ch1.sec1.thm1.eq3}
  dV(X(t))=dV_{1}(X(t))+dV_{2}(X(t))+dV_{3}(X(t)),
\end{equation}
where
\begin{eqnarray}
  dV_{1}(X(t)) &=& \frac{1}{S(t)}dS(t)-\frac{1}{2}\frac{1}{S^{2}(t)}(dS(t))^{2}\nonumber \\
   &=&\left[ \frac{B}{S(t)}-\beta \int^{h_{1}}_{t_{0}}f_{T_{1}}(s) e^{-\mu_{v} s}G(I(t-s))ds - \mu + \frac{\alpha}{S(t)} \int_{t_{0}}^{\infty}f_{T_{3}}(r)I(t-r)e^{-\mu r}dr \right.\nonumber\\
   &&\left.-\frac{1}{2}\sigma^{2}_{S}-\frac{1}{2}\sigma^{2}_{\beta}\left(\int^{h_{1}}_{t_{0}}f_{T_{1}}(s) e^{-\mu_{v} s}G(I(t-s))ds\right)^{2}\right]dt\nonumber\\
&&-\sigma_{S}dw_{S}(t)-\sigma_{\beta} \int^{h_{1}}_{t_{0}}f_{T_{1}}(s) e^{-\mu_{v} s}G(I(t-s))dsdw_{\beta}(t), \label{ch1.sec1.thm1.eq4}
\end{eqnarray}
\begin{eqnarray}
  dV_{2}(X(t)) &=& \frac{1}{E(t)}dE(t)-\frac{1}{2}\frac{1}{E^{2}(t)}(dE(t))^{2} \nonumber\\
  &=& \left[ \beta \frac{S(t)}{E(t)}\int^{h_{1}}_{t_{0}}f_{T_{1}}(s) e^{-\mu_{v} s}G(I(t-s))ds - \mu \right.\nonumber\\
&&\left.-\beta\frac{1}{E(t)} \int_{t_{0}}^{h_{2}}f_{T_{2}}(u)S(t-u)\int^{h_{1}}_{t_{0}}f_{T_{1}}(s) e^{-\mu s-\mu u}G(I(t-s-u))dsdu \right.\nonumber\\
&&\left.-\frac{1}{2}\sigma^{2}_{E}-\frac{1}{2}\sigma^{2}_{\beta}\frac{S^{2}(t)}{E^{2}(t)}\left(\int^{h_{1}}_{t_{0}}f_{T_{1}}(s) e^{-\mu_{v} s}G(I(t-s))ds\right)^{2}\right.\nonumber\\
&&\left.-\frac{1}{2}\sigma^{2}_{\beta}\frac{1}{E^{2}(t)}\left(\int_{t_{0}}^{h_{2}}f_{T_{2}}(u)S(t-u)\int^{h_{1}}_{t_{0}}f_{T_{1}}(s) e^{-\mu_{v} s-\mu u}G(I(t-s-u))dsdu \right)^{2}\right]dt\nonumber\\
&&-\sigma_{E}dw_{E}(t)+\sigma_{\beta} \frac{S(t)}{E(t)}\int^{h_{1}}_{t_{0}}f_{T_{1}}(s) e^{-\mu_{v} s}G(I(t-s))dsdw_{\beta}(t)\nonumber\\
&&-\sigma_{\beta}\frac{1}{E(t)} \int_{t_{0}}^{h_{2}}f_{T_{2}}(u)S(t-u)\int^{h_{1}}_{t_{0}}f_{T_{1}}(s) e^{-\mu_{v} s-\mu u}G(I(t-s-u))dsdudw_{\beta}(t),\nonumber\\
\label{ch1.sec1.thm1.eq5}
\end{eqnarray}
and
\begin{eqnarray}
  dV_{3}(X(t)) &=& \frac{1}{I(t)}dI(t)-\frac{1}{2}\frac{1}{I^{2}(t)}(dI(t))^{2}\nonumber \\
  &=& \left[\beta \frac{1}{I(t)}\int_{t_{0}}^{h_{2}}f_{T_{2}}(u)S(t-u)\int^{h_{1}}_{t_{0}}f_{T_{1}}(s) e^{-\mu_{v} s-\mu u}G(I(t-s-u))dsdu- (\mu +d+ \alpha)\right.  \nonumber\\
  &&\left.-\frac{1}{2}\sigma^{2}_{I}-\frac{1}{2}\sigma^{2}_{\beta}\left(\int_{t_{0}}^{h_{2}}f_{T_{2}}(u)S(t-u)\int^{h_{1}}_{t_{0}}f_{T_{1}}(s) e^{-\mu_{v} s-\mu u}G(I(t-s-u))dsdu \right)^{2}\right]dt\nonumber\\
&&-\sigma_{I}dw_{I}(t)+\sigma_{\beta}\frac{1}{I(t)} \int_{t_{0}}^{h_{2}}f_{T_{2}}(u)S(t-u)\int^{h_{1}}_{t_{0}}f_{T_{1}}(s) e^{-\mu_{v} s-\mu u}G(I(t-s-u))dsdudw_{\beta}(t)\nonumber\\
&&\label{ch1.sec1.thm1.eq6}
\end{eqnarray}
It follows from (\ref{ch1.sec1.thm1.eq3})-(\ref{ch1.sec1.thm1.eq6}) that for $t<\tau_{+}(t)$,
\begin{eqnarray}
  V(X(t))-V(X(t_{0})) &\geq& \int^{t}_{t_{0}}\left[-\beta \int^{h_{1}}_{t_{0}}f_{T_{1}}(s) e^{-\mu_{v} s}G(I(\xi-s))ds-\frac{1}{2}\sigma^{2}_{S}\right.\nonumber\\
   &&\left.-\frac{1}{2}\sigma^{2}_{\beta}\left(\int^{h_{1}}_{t_{0}}f_{T_{1}}(s) e^{-\mu_{v} s}G(I(\xi-s))ds\right)^{2}\right]d\xi\nonumber\\
   &&+ \int_{t}^{t_{0}}\left[-\beta\frac{1}{E(\xi)} \int_{t_{0}}^{h_{2}}f_{T_{2}}(u)S(\xi-u)\int^{h_{1}}_{t_{0}}f_{T_{1}}(s) e^{-\mu_{v} s-\mu u}G(I(\xi-s-u))dsdu
\right.\nonumber\\
&&\left.-\frac{1}{2}\sigma^{2}_{E}-\frac{1}{2}\sigma^{2}_{\beta}\frac{S^{2}(\xi)}{E^{2}(\xi)}\left(\int^{h_{1}}_{t_{0}}f_{T_{1}}(s) e^{-\mu_{v} s}G(I(\xi-s))ds\right)^{2}\right.\nonumber\\
&&\left.-\frac{1}{2}\sigma^{2}_{\beta}\frac{1}{E^{2}(\xi)}\left(\int_{t_{0}}^{h_{2}}f_{T_{2}}(u)S(\xi-u)\int^{h_{1}}_{t_{0}}f_{T_{1}}(s) e^{-\mu_{v} s-\mu u}G(I(\xi-s-u))dsdu \right)^{2}\right]d\xi\nonumber\\
   &&+ \int_{t}^{t_{0}}\left[- (3\mu +d+ \alpha)-\frac{1}{2}\sigma^{2}_{I}\right.  \nonumber\\
  &&\left.-\frac{1}{2}\sigma^{2}_{\beta}\left(\int_{t_{0}}^{h_{2}}f_{T_{2}}(u)S(\xi-u)\int^{h_{1}}_{t_{0}}f_{T_{1}}(s) e^{-\mu_{v} s-\mu u}G(I(\xi-s-u))dsdu \right)^{2}\right]d\xi\nonumber\\
&&+\int_{t}^{t_{0}}\left[-\sigma_{S}dw_{S}(\xi)-\sigma_{\beta} \int^{h_{1}}_{t_{0}}f_{T_{1}}(s) e^{-\mu_{v} s}G(I(\xi-s))dsdw_{\beta}(\xi)\right]\nonumber \\
  &&+\int_{t}^{t_{0}}\left[-\sigma_{E}dw_{E}(\xi)+\sigma_{\beta} \frac{S(\xi)}{E(\xi)}\int^{h_{1}}_{t_{0}}f_{T_{1}}(s) e^{-\mu_{v} s}G(I(\xi-s))dsdw_{\beta}(\xi)\right]\nonumber\\
&&-\int_{t}^{t_{0}}\left[\sigma_{\beta}\frac{1}{E(\xi)} \int_{t_{0}}^{h_{2}}f_{T_{2}}(u)S(\xi-u)\int^{h_{1}}_{t_{0}}f_{T_{1}}(s) e^{-\mu_{v} s-\mu u}G(I(\xi-s-u))dsdudw_{\beta}(\xi)\right]\nonumber\\
&&+\int_{t_{0}}^{t}\left[-\sigma_{I}dw_{I}(\xi)\right.\nonumber\\
&&\left.+\sigma_{\beta}\frac{1}{I(\xi)} \int_{t_{0}}^{h_{2}}f_{T_{2}}(u)S(\xi-u)\int^{h_{1}}_{t_{0}}f_{T_{1}}(s) e^{-\mu_{v} s-\mu u}G(I(\xi-s-u))dsdudw_{\beta}(\xi)\right].\nonumber\\
&&\label{ch1.sec1.thm1.eq7}
\end{eqnarray}
Taking the limit on (\ref{ch1.sec1.thm1.eq7}) as $t\rightarrow \tau_{+}(t)$, it follows from (\ref{ch1.sec1.thm1.eq1})-(\ref{ch1.sec1.thm1.eq2}) that the left-hand side $V(X(t))-V(X(t_{0}))\leq -\infty$. This contradicts the finiteness of the right-handside of the inequality (\ref{ch1.sec1.thm1.eq7}). Hence $\tau_{+}(t)=\tau_{e}(w)$ a.s., that is, $X(t,w)\in D(\tau_{e})$,   whenever  $\sigma_{i}=0$, $i\in \{S, E, I\}$ and $\sigma_{\beta}\geq 0$, and $X(t,w)\in \mathbb{R}^{4}_{+}$, whenever  $\sigma_{i}>0$, $i\in \{S, E, I\}$ and $\sigma_{\beta}\geq 0$.

The following shows that $\tau_{e}(w)=\infty$. Let $k>0$ be a positive integer such that $||\vec{\varphi}||_{1}\leq k$, where $\vec{\varphi}=\left(\varphi_{1}(t),\varphi_{2}(t), \varphi_{3}(t)\right), t\in (-\infty,t_{0}]$ defined in (\ref{ch1.sec0.eq12}), and $||.||_{1}$ is the $p-sum$ norm defined on $\mathbb{R}^{3}$, when $p=1$. Define the stopping time
\begin{equation}\label{ch1.sec1.thm1.eq8}
\left\{
\begin{array}{ll}
\tau_{k}=sup\{t\in [t_{0},\tau_{e}): ||X(s)||_{1}=S(s)+E(s)+I(s)\leq k, s\in[t_{0},t] \}\\
\tau_{k}(t)=\min(t,\tau_{k}).
\end{array}
\right.
\end{equation}
It is easy to see that as $k\rightarrow \infty$, $\tau_{k}$ increases. Set $\lim_{k\rightarrow \infty}\tau_{k}(t)=\tau_{\infty}$. Then it follows that $\tau_{\infty}\leq \tau_{e}$ a.s.
 We show in the following that: (1.) $\tau_{e}=\tau_{\infty}\quad a.s.\Leftrightarrow P(\tau_{e}\neq \tau_{\infty})=0$, (2.)  $\tau_{\infty}=\infty\quad a.s.\Leftrightarrow P(\tau_{\infty}=\infty)=1$.

Suppose on the contrary that $P(\tau_{\infty}<\tau_{e})>0$. Let $w\in \{\tau_{\infty}<\tau_{e}\}$ and $t\leq \tau_{\infty}$.
 Define
\begin{equation}\label{ch1.sec1.thm1.eq9}
\left\{
\begin{array}{ll}
\hat{V}_{1}(X(t))=e^{\mu t}(S(t)+E(t)+I(t)),\\
\forall t\leq\tau_{k}(t).
\end{array}
\right.
\end{equation}
The Ito-Doob differential $d\hat{V}_{1}$ of (\ref{ch1.sec1.thm1.eq9}) with respect to the system (\ref{ch1.sec0.eq8})-(\ref{ch1.sec0.eq11}) is given as follows:
\begin{eqnarray}
 d\hat{V}_{1} &=& \mu e^{\mu t}(S(t)+E(t)+I(t)) dt + e^{\mu t}(dS(t)+dE(t)+dI(t))  \\
   &=& e^{\mu t}\left[B+\alpha \int_{t_{0}}^{\infty}f_{T_{3}}(r)I(t-r)e^{-\mu r}dr-(\alpha + d)I(t)\right]dt\nonumber\\
   &&-\sigma_{S}e^{\mu t}S(t)dw_{S}(t)-\sigma_{E}e^{\mu t}E(t)dw_{E}(t)-\sigma_{I}e^{\mu t}I(t)dw_{I}(t)\label{ch1.sec1.thm1.eq10}
\end{eqnarray}
Integrating (\ref{ch1.sec1.thm1.eq9}) over the interval $[t_{0}, \tau]$, and applying some algebraic manipulations and  simplifications  it follows that
\begin{eqnarray}
  V_{1}(X(\tau)) &=& V_{1}(X(t_{0}))+\frac{B}{\mu}\left(e^{\mu \tau}-e^{\mu t_{0}}\right)\nonumber\\
  &&+\int_{t_{0}}^{\infty}f_{T_{3}}(r)e^{-\mu r}\left(\int_{t_{0}-r}^{t_{0}}\alpha I(\xi)d\xi-\int_{\tau-r}^{\tau}\alpha I(\xi)d\xi\right)dr-\int_{t_{0}}^{\tau}d I(\xi)d\xi \nonumber\\
  &&+\int^{\tau}_{t_{0}}\left[-\sigma_{S}e^{\mu \xi}S(\xi)dw_{S}(\xi)-\sigma_{E}e^{\mu \xi}E(\xi)dw_{E}(\xi)-\sigma_{I}e^{\mu \xi}I(\xi)dw_{I}(\xi)\right]\label{ch1.sec1.thm1.eq11}
\end{eqnarray}
Removing negative terms from (\ref{ch1.sec1.thm1.eq11}), it implies from (\ref{ch1.sec0.eq12}) that
\begin{eqnarray}
  V_{1}(X(\tau)) &\leq& V_{1}(X(t_{0}))+\frac{B}{\mu}e^{\mu \tau}\nonumber\\
  &&+\int_{t_{0}}^{\infty}f_{T_{3}}(r)e^{-\mu r}\left(\int_{t_{0}-r}^{t_{0}}\alpha \varphi_{3}(\xi)d\xi\right)dr \nonumber\\
  &&+\int^{\tau}_{t_{0}}\left[-\sigma_{S}e^{\mu \xi}S(\xi)dw_{S}(\xi)-\sigma_{E}e^{\mu \xi}E(\xi)dw_{E}(\xi)-\sigma_{I}e^{\mu \xi}I(\xi)dw_{I}(\xi)\right]\label{ch1.sec1.thm1.eq12}
\end{eqnarray}
But from (\ref{ch1.sec1.thm1.eq9}) it is easy to see that for $\forall t\leq\tau_{k}(t)$,
\begin{equation}\label{ch1.sec1.thm1.eq12a}
  ||X(t)||_{1}=S(t)+E(t)+I(t)\leq V(X(t)).
\end{equation}
 Thus setting $\tau=\tau_{k}(t)$, then it follows from
(\ref{ch1.sec1.thm1.eq8}), (\ref{ch1.sec1.thm1.eq12}) and  (\ref{ch1.sec1.thm1.eq12a}) that
\begin{equation}\label{ch1.sec1.thm1.eq13}
  k=||X(\tau_{k}(t))||_{1}\leq V_{1}(X(\tau_{k}(t)))
\end{equation}
Taking the limit on (\ref{ch1.sec1.thm1.eq13}) as $k\rightarrow \infty$ leads to a contradiction because the left-hand-side of the inequality (\ref{ch1.sec1.thm1.eq13}) is infinite, but following the right-hand-side  from (\ref{ch1.sec1.thm1.eq12}) leads to a finite value. Hence $\tau_{e}=\tau_{\infty}$ a.s. The following shows that $\tau_{e}=\tau_{\infty}=\infty$ a.s.

  Let $\ w\in \{\tau_{e}<\infty\}$. It follows from (\ref{ch1.sec1.thm1.eq11})-(\ref{ch1.sec1.thm1.eq12}) that
  \begin{eqnarray}
  I_{\{\tau_{e}<\infty\}}V_{1}(X(\tau)) &\leq& I_{\{\tau_{e}<\infty\}}V_{1}(X(t_{0}))+I_{\{\tau_{e}<\infty\}}\frac{B}{\mu}e^{\mu \tau}\nonumber\\
  &&+I_{\{\tau_{e}<\infty\}}\int_{t_{0}}^{\infty}f_{T_{3}}(r)e^{-\mu r}\left(\int_{t_{0}-r}^{t_{0}}\alpha \varphi_{3}(\xi)d\xi\right)dr\nonumber\\
  &&+I_{\{\tau_{e}<\infty\}}\int^{\tau}_{t_{0}}\left[-\sigma_{S}e^{\mu \xi}S(\xi)dw_{S}(\xi)-\sigma_{E}e^{\mu \xi}E(\xi)dw_{E}(\xi)-\sigma_{I}e^{\mu \xi}I(\xi)dw_{I}(\xi)\right].
  \nonumber\\
  \label{ch1.sec1.thm1.eq14}
\end{eqnarray}
Suppose $\tau=\tau_{k}(t)\wedge T$, where $ T>0$ is arbitrary, then taking the expected value of (\ref{ch1.sec1.thm1.eq14}) follows that
\begin{equation}\label{ch1.sec1.thm1.eq14a}
  E(I_{\{\tau_{e}<\infty\}}V_{1}(X(\tau_{k}(t)\wedge T))) \leq V_{1}(X(t_{0}))+\frac{B}{\mu}e^{\mu T}
\end{equation}
But from (\ref{ch1.sec1.thm1.eq12a}) it is easy to see that
\begin{equation}\label{ch1.sec1.thm1.eq15}
 I_{\{\tau_{e}<\infty,\tau_{k}(t)\leq T\}}||X(\tau_{k}(t))||_{1}\leq I_{\{\tau_{e}<\infty\}}V_{1}(X(\tau_{k}(t)\wedge T))
\end{equation}
It follows from (\ref{ch1.sec1.thm1.eq14})-(\ref{ch1.sec1.thm1.eq15}) and
   (\ref{ch1.sec1.thm1.eq8}) that
 \begin{eqnarray}
 P(\{\tau_{e}<\infty,\tau_{k}(t)\leq T\})k&=&E\left[I_{\{\tau_{e}<\infty,\tau_{k}(t)\leq T\}}||X(\tau_{k}(t))||_{1}\right]\nonumber\\
 &\leq& E\left[I_{\{\tau_{e}<\infty\}}V_{1}(X(\tau_{k}(t)\wedge T))\right]\nonumber\\
 &\leq& V_{1}(X(t_{0}))+\frac{B}{\mu}e^{\mu T}.
\label{ch1.sec1.thm1.eq16}
 \end{eqnarray}
  It follows immediately from (\ref{ch1.sec1.thm1.eq16}) that
 $P(\{\tau_{e}<\infty,\tau_{\infty}\leq T\})\rightarrow 0$ as $k\rightarrow \infty$. Furthermore, since $T<\infty$ is arbitrary, we conclude that $P(\{\tau_{e}<\infty,\tau_{\infty}< \infty\})= 0$.
Finally,  by the total probability principle,
 \begin{eqnarray}
 P(\{\tau_{e}<\infty\})&=&P(\{\tau_{e}<\infty,\tau_{\infty}=\infty\})+P(\{\tau_{e}<\infty,\tau_{\infty}<\infty\})\nonumber\\
 &\leq&P(\{\tau_{e}\neq\tau_{\infty}\})+P(\{\tau_{e}<\infty,\tau_{\infty}<\infty\})\nonumber\\
 &=&0.\label{ch1.sec1.thm1.eq17}
 \end{eqnarray}
 Thus from (\ref{ch1.sec1.thm1.eq17}), $\tau_{e}=\tau_{\infty}=\infty$ a.s.. In addition, $X(t)\in D(\infty)$, whenever  $\sigma_{i}=0$, $i\in \{S, E, I\}$ and $\sigma_{\beta}\geq 0$, and $X(t,w)\in \mathbb{R}^{4}_{+}$, whenever  $\sigma_{i}>0$, $i\in \{S, E, I\}$ and $\sigma_{\beta}\geq 0$.
\begin{rem}\label{ch1.sec0.remark1}
Theorem~\ref{ch1.sec1.thm1} and Lemma~\ref{ch1.sec1.lemma1} signify that the stochastic system (\ref{ch1.sec0.eq8})-(\ref{ch1.sec0.eq11}) has a unique positive solution  $Y(t)\in \mathbb{R}^{4}_{+}$ globally for all $t\in (-\infty, \infty)$. Furthermore, it follows that a positive solution of the stochastic system that starts in the closed ball centered at the origin with a radius of $\frac{B}{\mu}$, $D(\infty)=\bar{B}^{(-\infty, \infty)}_{\mathbb{R}^{4}_{+},}\left(0,\frac{B}{\mu}\right)$, will continue to oscillate and remain bounded in the closed ball for all time $t\geq t_{0}$, whenever the intensities of the independent white noise processes in the system satisfy  $\sigma_{i}=0$, $i\in \{S, E, I\}$ and $\sigma_{\beta}\geq 0$. Hence, the set $D(\infty)=\bar{B}^{(-\infty, \infty)}_{\mathbb{R}^{4}_{+},}\left(0,\frac{B}{\mu}\right)$ is a positive self-invariant set for the stochastic system (\ref{ch1.sec0.eq8})-(\ref{ch1.sec0.eq11}). In the case where the intensities of the independent white noise processes in the system satisfy  $\sigma_{i}>0$, $i\in \{S, E, I\}$ and $\sigma_{\beta}\geq 0$, the solution are positive and unique, and continue to oscillate in the unbounded space of positive real numbers $\mathbb{R}^{4}_{+}$. In other words, the positive solutions of the system are bounded, whenever $\sigma_{i}=0$, $i\in \{S, E, I\}$ and $\sigma_{\beta}\geq 0$ and unbounded,  whenever $\sigma_{i}>0$, $i\in \{S, E, I\}$ and $\sigma_{\beta}\geq 0$.

The implication of this result to the disease dynamics represented by (\ref{ch1.sec0.eq8})-(\ref{ch1.sec0.eq11}) is that the occurrence of noise exclusively from the disease transmission rate allows a controlled situation for the disease dynamics, since the positive solutions exist within a positive self invariant space. The additional source of variability from the natural death rate can lead to more complex and uncontrolled situations for the disease dynamics, since it is obvious that the intensities of the white noise processes from the natural death rates of the different states in the system are driving the positive solutions of the system unbounded. Some examples of uncontrolled disease situations that can occur when the positive solutions are unbounded include:-  (1) extinction of the population, (2) failure to find an infection-free steady population state, wherein the disease be controlled by bringing the population into that state, and (3)  a sudden significant random flip of a given state such as the infectious state from a low to high value, or vice versa over a short time interval etc.  These facts become more apparent in the subsequent sections where conditions for disease eradication are derived.
\end{rem}
 \section{Extinction of disease with noise from both disease transmission and natural death rates \label{ch1.sec2a}}
The extinction of the vector-borne disease from the population described by the stochastic epidemic dynamic model (\ref{ch1.sec0.eq8})-(\ref{ch1.sec0.eq11}) is exhibited in this section. Noting that (\ref{ch1.sec0.eq9}) and (\ref{ch1.sec0.eq11}) decouple from the system (\ref{ch1.sec0.eq8})-(\ref{ch1.sec0.eq11}), it follows that the two other equations for $S(t)$ and $I(t)$ in  (\ref{ch1.sec0.eq8}) and (\ref{ch1.sec0.eq10}), respectively, depend only on the states $(S(t), I(t))$. Therefore, it suffices to show extinction of the disease from the population by showing the extinction of the infectious population $I(t)$.

Recall Theorem~\ref{ch1.sec1.thm1}(b), asserts that the system  (\ref{ch1.sec0.eq8})-(\ref{ch1.sec0.eq11}) has a unique solution process $\{Y(t),t\geq t_{0}\}$ with positive solution paths for the malaria dynamics. Furthermore, all paths that start in $\mathbb{R}^{4}_{+}$ continue to oscillate in the space $\mathbb{R}^{4}_{+}$. Moreover, it was remarked  in Remark~\ref{ch1.sec0.remark1} that the solution paths are potentially liable to become unbounded in the space $\mathbb{R}^{4}_{+}$, whenever the strength of the noise  from the natural death rates is strong.

In this section, the threshold conditions for the intensities  $\sigma_{i},i\in\{S, E, I, R\}$, and other parameters of the disease dynamics in (\ref{ch1.sec0.eq8})-(\ref{ch1.sec0.eq11}), which are sufficient for extinction are presented. Recall \cite{zhien,xin}, the following definition of the extinction of a species denoted by the process $Z(t), t\geq t_{0}$ in a stochastic dynamic system:
\begin{defn}\label{ch1.sec2a.defn1}
\item{(1.)} $Z(t)$ is said to  be extinct if $\lim_{t\rightarrow \infty}{Z(t)}=0$, a.s.
\item{(2.)} $Z(t)$ is said to  be stable in the mean if $\lim_{t\rightarrow \infty}{\frac{1}{t}\int^{t}_{t_{0}}Z(s)}ds=c>0$, a.s.
\item{(3.)} $Z(t)$ is said to  be strongly persistent in the mean if $\liminf_{t\rightarrow \infty}{\frac{1}{t}\int^{t}_{t_{0}}Z(s)}ds>0$, a.s.
\end{defn}
That is, the species is extinct if every path for the process $Z(t), t\geq t_{0}$ converges to zero with probability one, and stable in the mean, if every path converges in the mean asymptotically to a constant, with probability one. Note that if the species is stable in the mean, then it is also strongly persistent, but the converse is not always true.

\begin{figure}[H]
\begin{center}
\includegraphics[height=8cm]{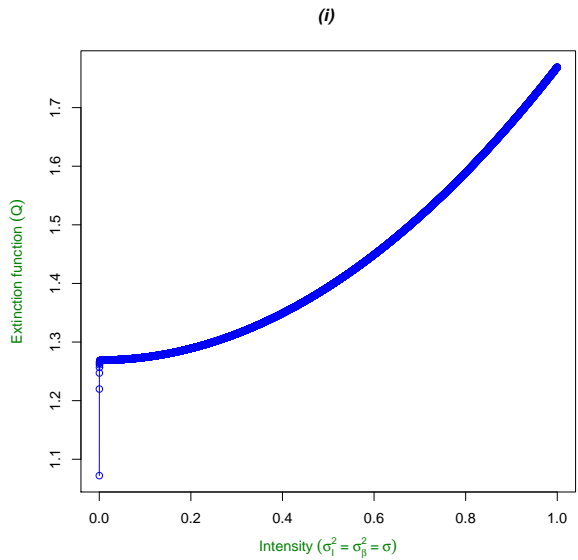}
\caption{(i) shows the behavior of the extinction function (rate) $Q$ when the values of $\sigma^{2}_{\beta}=\sigma^{2}_{I}=\sigma^{2}$ continuously increase over the range $[0,1]$. It can be seen that larger values of $\sigma^{2}_{\beta}=\sigma^{2}_{I}=\sigma^{2}$ correspond to larger values of $Q$. This suggests that as the values of $\sigma^{2}_{\beta}=\sigma^{2}_{I}=\sigma^{2}$ continuously increase, the disease becomes extinct at a faster rate. }\label{ch1.sec2a.rem1.figure1}
\end{center}
\end{figure}

Also, the following lemma from \cite{mao} known as the exponential martingale inequality will be used to establish the extinction results, whenever Theorem~\ref{ch1.sec1.thm1}(b) holds.
\begin{lemma}\label{ch1.sec2a.lem1}
Let $g=(g_{1},g_{2},\ldots, g_{m})\in L^{2}(\mathbb{R}_{+},\mathbb{R}^{1\times m} )$, and $T, c, \theta$ be any positive numbers. Then
\begin{equation}\label{ch1.sec2a.lemma1.eq1}
  P\left( \sup_{0\leq t\leq T}{\left[\int^{t}_{0}g(s)dB(s)-\frac{c}{2}\int^{t}_{0}|g(s)|^{2}ds\right]>\theta}\right)\leq e^{-c\theta}.
\end{equation}
\end{lemma}
Proof: See \cite{mao}
\begin{thm}\label{ch1.sec2a.thm1}
Let $\sigma_{i}>0,\forall i\in \{S, E, I, R,\beta\}$, and let $(S(t), I(t))$ be the solution of the decoupled system (\ref{ch1.sec0.eq8}) and (\ref{ch1.sec0.eq10}) with initial conditions in (\ref{ch1.sec0.eq12}) and (\ref{ch1.sec0.eq13}), that satisfies  Theorem~\ref{ch1.sec1.thm1}(b). Suppose further that  the following relationship holds
\begin{equation}\label{ch1.sec2a.thm1.eq1}
  \sigma^{2}_{\beta}> \frac{\beta^{2}}{2(\mu+ d+ \alpha)+\sigma^{2}_{I}}.
\end{equation}
Then it follows that the solution of the decoupled system (\ref{ch1.sec0.eq8}) and (\ref{ch1.sec0.eq10}) satisfies
\begin{equation}\label{ch1.sec2a.thm1.eq2}
  \limsup_{t\rightarrow \infty}{\frac{1}{t}\log{(I(t))}}\leq \frac{\beta^{2}}{2\sigma^{2}_{\beta}}-(\mu + d+ \alpha + \frac{1}{2}\sigma^{2}_{I})<0 \quad a.s.
\end{equation}
That is, $I(t)$ tends to zero exponentially almost surely. In other words, the infectious population is extinct and the disease dies out with probability one.
\end{thm}
Proof:\\
The differential operator $dV$ applied to the function
\begin{equation}\label{ch1.sec2a.thm1.proof.eq1}
  V(t)=\log{I(t)},
\end{equation}
with respect to the system (\ref{ch1.sec0.eq8}) and (\ref{ch1.sec0.eq10}) leads to the following
\begin{equation}\label{ch1.sec2a.thm1.proof.eq2}
  dV(t)=f(S, I)dt -\sigma_{I}dw_{I}(t) + \sigma_{\beta} \int_{t_{0}}^{h_{2}}\int^{h_{1}}_{t_{0}}f_{T_{2}}(u)f_{T_{1}}(s) e^{-(\mu_{v} s+\mu u)}S(t-u)\frac{G(I(t-s-u))}{I(t)}dsdudw_{\beta}(t),
\end{equation}
where,
\begin{eqnarray}
  f(S, I)&=&\beta\int_{t_{0}}^{h_{2}}f_{T_{2}}(u)\int^{h_{1}}_{t_{0}}f_{T_{1}}(s) e^{-(\mu_{v} s+\mu u)}S(t-u)\frac{G(I(t-s-u))}{I(t)}dsdu-(\mu+ d+ \alpha+\frac{1}{2}\sigma^{2}_{I})\nonumber\\
  &&-\frac{1}{2}\sigma^{2}_{\beta}\left(\int_{t_{0}}^{h_{2}}f_{T_{2}}(u)\int^{h_{1}}_{t_{0}}f_{T_{1}}(s) e^{-(\mu_{v} s+\mu u)}S(t-u)\frac{G(I(t-s-u))}{I(t)}dsdu\right)^{2}.\label{ch1.sec2a.thm1.proof.eq3}
\end{eqnarray}
  Define the following
\begin{eqnarray}
Z(t)&=&\int_{t_{0}}^{h_{2}}\int^{h_{1}}_{t_{0}}f_{T_{2}}(u)f_{T_{1}}(s) e^{-(\mu_{v} s+\mu u)}S(t-u)\frac{G(I(t-s-u))}{I(t)}dsdu\label{ch1.sec2a.thm1.proof.eq4a}\\
  M_{1}(t)&=&\int^{t}_{t_{0}}\sigma_{I}dw_{I}(\xi)=\sigma_{I}(w_{I}(t)-w_{I}(t_{0})),\label{ch1.sec2a.thm1.proof.eq4}\\
  M_{2}(t)&=&\sigma_{\beta}\int^{t}_{t_{0}} \int_{t_{0}}^{h_{2}}\int^{h_{1}}_{t_{0}}f_{T_{2}}(u)f_{T_{1}}(s) e^{-(\mu_{v} s+\mu u)}S(\xi-u)\frac{G(I(\xi-s-u))}{I(\xi)}dsdudw_{\beta}(\xi)\nonumber\\
  &=&\sigma_{\beta}\int^{t}_{t_{0}}Z(\xi)dw_{\beta}(\xi).\label{ch1.sec2a.thm1.proof.eq5}
\end{eqnarray}
It is easy to see that $M_{2}(t)$ is a local martingale with a quadratic variation given by
\begin{equation}\label{ch1.sec2a.thm1.proof.eq5a}
  <M_{2}(t), M_{2}(t)>=\sigma^{2}_{\beta}\int^{t}_{t_{0}}Z^{2}(\xi)d\xi.
\end{equation}
Furthermore, utilizing the exponential martingale inequality in Lemma~\ref{ch1.sec2a.lem1}, it follows that for any random integer $k\equiv k(w), w\in \Omega$, and constant $0<c<1$, the probability of the event $A_{k}$ defined below
\begin{equation}\label{ch1.sec2a.thm1.proof.eq5b}
  P(A_{k})=P\left(\left\{w\in \Omega: \sup_{t_{0}\leq t\leq k(w)}{\left[M_{2}(t)-\frac{c}{2} <M_{2}(t), M_{2}(t)>\right]>\frac{2}{c}\log{k(w)}}\right\}\right)\leq \frac{1}{(k)^{2}}.
\end{equation}
 The sequence of events $\left\{A_{k}\right\}_{k=0}^{\infty}$  satisfies  $\sum_{k}{P(A_{k})}<\infty $, and consequently by Borel-Cantelli Lemma\cite{mao}, there exists a random integer $k_{0}\equiv k_{0}(w)>t_{0}$ such that
 \begin{equation}\label{ch1.sec2a.thm1.proof.eq5c}
   \sup_{t_{0}\leq t\leq k}{\left[M_{2}(t)-\frac{c}{2} <M_{2}(t), M_{2}(t)>\right]\leq\frac{2}{c}\log{k}},\quad a.s.,
 \end{equation}
 whenever $k>k_{0}$. And (\ref{ch1.sec2a.thm1.proof.eq5c}) further leads to
 \begin{equation}\label{ch1.sec2a.thm1.proof.eq5d}
   M_{2}(t)\leq \frac{1}{2}c\sigma^{2}_{\beta}\int^{t}_{t_{0}}Z^{2}(\xi)d\xi + \frac{2}{c}\log{k}, \forall t\in [t_{0}, k].
 \end{equation}
 Now, integrating both sides of  (\ref{ch1.sec2a.thm1.proof.eq2}) over the interval $[t_{0},t]$, it follows from (\ref{ch1.sec2a.thm1.proof.eq3})-(\ref{ch1.sec2a.thm1.proof.eq5d}) and some algebraic manipulations and simplifications that for any $t\in [t_{0},k]$,
 \begin{eqnarray}
   \log{I(t)}&\leq& \log{I(t_{0})}+\int_{t_{0}}^{t}\left[\frac{\beta^{2}}{2\sigma^{2}_{\beta}(1-c)}-(\mu + d+ \alpha + \frac{1}{2}\sigma^{2}_{I})\right]d\xi+\frac{2}{c}\log{k}\nonumber\\
   &&-\frac{1}{2}\sigma^{2}_{\beta}(1-c)\int_{t_{0}}^{t}\left(Z(\xi)-\frac{\beta}{\sigma^{2}_{\beta}(1-c)}\right)^{2}d\xi-M_{1}(t).\label{ch1.sec2a.thm1.proof.eq5e}
 \end{eqnarray}
 Moreover, (\ref{ch1.sec2a.thm1.proof.eq5e}) simplifies to
 \begin{eqnarray}
   \log{I(t)}&\leq& \log{I(t_{0})}+\left[\frac{\beta^{2}}{2\sigma^{2}_{\beta}(1-c)}-(\mu + d+ \alpha + \frac{1}{2}\sigma^{2}_{I})\right](t-t_{0})+\frac{2}{c}\log{k}\nonumber\\
   &&-M_{1}(t).\label{ch1.sec2a.thm1.proof.eq5f}
 \end{eqnarray}
 Diving both sides of (\ref{ch1.sec2a.thm1.proof.eq5f}) by $t$, it follows that for $k-1\leq t\leq k$, one obtains the following inequality
  \begin{eqnarray}
   \frac{1}{t}\log{I(t)}&\leq& \frac{1}{t}\log{I(t_{0})}+\left[\frac{\beta^{2}}{2\sigma^{2}_{\beta}(1-c)}-(\mu + d+ \alpha + \frac{1}{2}\sigma^{2}_{I})\right](1-\frac{t_{0}}{t})+\frac{2}{c}\frac{\log{k}}{k-1}\nonumber\\
   &&-\frac{1}{t}M_{1}(t).\label{ch1.sec2a.thm1.proof.eq5g}
 \end{eqnarray}
 It follows further that for sufficiently large $k$ (i.e. $k\rightarrow \infty$), then $t\rightarrow\infty$, and consequently, taking the limit supremum of (\ref{ch1.sec2a.thm1.proof.eq5g}) as $t\rightarrow\infty$, it is easy to see that (\ref{ch1.sec2a.thm1.proof.eq5g}) reduces to
 \begin{eqnarray}
   \limsup_{t\rightarrow\infty}\frac{1}{t}\log{I(t)}&\leq& \left[\frac{\beta^{2}}{2\sigma^{2}_{\beta}(1-c)}-(\mu + d+ \alpha + \frac{1}{2}\sigma^{2}_{I})\right]
   -\limsup_{t\rightarrow\infty}\frac{1}{t}M_{1}(t).\label{ch1.sec2a.thm1.proof.eq5h}
 \end{eqnarray}
But, it is easy to see  from the strong law of large numbers for local martingales (see, e.g. \cite{mao}) that
\begin{equation}\label{ch1.sec2a.thm1.proof.eq6}
  \limsup_{t\rightarrow \infty}{\frac{1}{t}M_{1}(t)}=0, a.s.
\end{equation}
Consequently, (\ref{ch1.sec2a.thm1.proof.eq5g}) reduces to
 \begin{eqnarray}
   \limsup_{t\rightarrow\infty}\frac{1}{t}\log{I(t)}&\leq& \left[\frac{\beta^{2}}{2\sigma^{2}_{\beta}(1-c)}-(\mu + d+ \alpha + \frac{1}{2}\sigma^{2}_{I})\right].
   \label{ch1.sec2a.thm1.proof.eq7}
 \end{eqnarray}
 Thus, for $c$ infinitesimally small, that is, $c\rightarrow 0$,  (\ref{ch1.sec2a.thm1.eq2}) follows immediately from (\ref{ch1.sec2a.thm1.proof.eq7}).
\begin{rem}\label{ch1.sec2a.rem1}
Theorem~\ref{ch1.sec2a.thm1} and Theorem~\ref{ch1.sec1.thm1}[b] signify that when the intensities of the disease transmission and natural death rates, $\sigma_{\beta}$ and $\sigma_{I}$, respectively,   are positive, then all sample paths of the solution process $\{(S(t), I(t)),t\geq t_{0}\}$ of the decoupled system (\ref{ch1.sec0.eq9}) and (\ref{ch1.sec0.eq11}) that start  in $\mathbb{R}^{2}_{+}$ continue to oscillate in $\mathbb{R}^{2}_{+}$. Moreover, the sample paths of the infectious state $I(t), t\geq t_{0}$ of the solution process $\{(S(t), I(t)),t\geq t_{0}\}$ ultimately turn to zero exponentially, almost surely, whenever the intensities of the disease transmission and natural death rates, $\sigma_{\beta}$ and $\sigma_{I}$, respectively, are related as shown in (\ref{ch1.sec2a.thm1.eq1}). Furthermore, the sample Lyapunov exponent from (\ref{ch1.sec2a.thm1.eq2}) is estimated by the  term $Q$, expressed as a function of $\sigma_{\beta}$ and $\sigma_{I}$ as follows
\begin{equation}\label{ch1.sec2a.rem1.eq1}
  \limsup_{t\rightarrow \infty}{\frac{1}{t}\log{(I(t))}}\leq -Q(\sigma^{2}_{\beta},\sigma^{2}_{I}) \quad a.s.,
\end{equation}
where
\begin{equation}\label{ch1.sec2a.rem1.eq2}
 Q(\sigma^{2}_{\beta},\sigma^{2}_{I})=(\mu + d+ \alpha + \frac{1}{2}\sigma^{2}_{I})-\frac{\beta^{2}}{2\sigma^{2}_{\beta}}.
\end{equation}
  It follows from (\ref{ch1.sec2a.rem1.eq1})-(\ref{ch1.sec2a.rem1.eq2}) that when the condition (\ref{ch1.sec2a.thm1.eq1}) holds, then the infectious population $I(t)$ dies out exponentially, almost surely, whenever the function $Q$ in (\ref{ch1.sec2a.rem1.eq2}) is positive, that is, $Q>0$. In addition, the rate of the exponential decay of each sample path of the infectious population $I(t)$ is given by the  estimate $Q(\sigma^{2}_{\beta},\sigma^{2}_{I})$ of the sample Lyapunov exponent in (\ref{ch1.sec2a.rem1.eq1}).

   The function $Q$ can be used to evaluate the qualitative effects of the intensities $\sigma_{\beta}$ and $\sigma_{I}$ on the rate of extinction of the disease from the system. Indeed, observe that the function $Q$ increases monotonically with respect to continuous changes in each intensity $\sigma_{\beta}$ and $\sigma_{I}$. This observation suggests that larger values of the intensities $\sigma_{\beta}$ and $\sigma_{I}$, lead to larger values of $Q$, and consequently lead to a larger rate of  extinction of the disease from the population. Figure~\ref{ch1.sec2a.rem1.figure1} illustrates the behavior of the decay rate $Q$, as the intensities  $\sigma_{\beta}$ and $\sigma_{I}$ of the independent white noise processes in the system continuously increase in value.

   The following assumptions hold for the example exhibited in Figure~\ref{ch1.sec2a.rem1.figure1}:  (1) the intensities of the random fluctuations in the disease transmission rate and natural death rate of infectious individuals, $\sigma_{\beta}$ and $\sigma_{I}$, respectively, continuously change equally, that is, $\sigma_{\beta}=\sigma_{I}$ , (2) the other parameters of the  system (\ref{ch1.sec0.eq8})-(\ref{ch1.sec0.eq11}) are selected conveniently as follows: the expected effective disease transmission rate $\beta=6.277E-5$, recovery rate $\alpha=0.55067$, average natural death rate of human beings  $\mu=0.6$, and disease related death rate $d=0.11838$.
\end{rem}
\section{Extinction  and stability of equilibrium with noise from disease transmission rate}\label{ch1.sec2b}
Recall Theorem~\ref{ch1.sec1.thm1}(a), asserts that the system  (\ref{ch1.sec0.eq8})-(\ref{ch1.sec0.eq11}) has a unique solution process $\{Y(t),t\geq t_{0}\}$ with positive solution paths for the malaria dynamics. In addition, all paths that start in $D(\infty)$ continue to oscillate in the space $D(\infty)$. Also, it was remarked earlier in Remark~\ref{ch1.sec0.remark1} that with noise in the disease dynamics exclusively from the disease transmission rate, that is, $\sigma_{\beta}>0$ and $\sigma_{i}=0, i\in \{S, E, I, R\}$, the unique solution paths are relatively more "well-behaved" with lesser tendency to  drift outside of the self-invariant space $D(\infty)$,  regardless whether the noise in the disease dynamics from the disease transmission rate is strong.

 In this section, it will be shown that the conditions for the extinction of the  disease from the population have no bearings on the intensity $\sigma_{\beta}$ of the noise in the disease transmission rate. In fact, it will is shown that the extinction of the disease from the population depends only on the  basic reproduction number $R^{*}_{0}$ in (\ref{ch1.sec2.lemma2a.corrolary1.eq4}) and (\ref{ch1.sec2.theorem1.corollary1.eq3}), or on the survival probability rate of the malaria parasites.
The following lemmas will be used to establish the extinction results, whenever Theorem~\ref{ch1.sec1.thm1}(a) holds.
\begin{lemma}\label{ch1.sec2b.lemma1}
Suppose Theorem~\ref{ch1.sec1.thm1}(a) holds, then the unique solution process $Y(t)\in D(\infty),t\geq t_{0}$ of the stochastic system (\ref{ch1.sec0.eq8})-(\ref{ch1.sec0.eq11}) also lies in the space
\begin{equation}\label{ch1.sec2b.lemma1.eq1}
  D^{expl}(\infty)=\left\{Y(t)\in \mathbb{R}^{4}_{+}:\frac{B}{\mu+d}\leq N(t)=S(t)+ E(t)+ I(t)+ R(t)\leq \frac{B}{\mu}, \forall t\in (-\infty, \infty) \right\},
\end{equation}
where $D^{expl}(\infty)\subset D(\infty)$.  Moreover, the space $D^{expl}(\infty)$ is also self-invariant with respect to the stochastic system (\ref{ch1.sec0.eq8})-(\ref{ch1.sec0.eq11}).
\end{lemma}
Proof:\\
Suppose Theorem~\ref{ch1.sec1.thm1}(a) holds, then it follows from (\ref{ch1.sec1.lemma1.eq2}) that the total population $N(t)=S(t)+E(t)+I(t)+R(t)$ satisfies the following inequality
\begin{equation}\label{ch1.sec2b.lemma1.proof.eq1}
[B-(\mu+d)N(t)]dt\leq dN(t)\leq [B-(\mu)N(t)]dt.
\end{equation}
It is easy to see from (\ref{ch1.sec2b.lemma1.proof.eq1}) that
\begin{equation}\label{ch1.sec2b.lemma1.proof.eq2}
\frac{B}{\mu+d}\leq \liminf_{t\rightarrow \infty}N(t)\leq \limsup_{t\rightarrow \infty}N(t)\leq \frac{B}{\mu},
\end{equation}
and (\ref{ch1.sec2b.lemma1.eq1}) follows immediately.
\begin{lemma}\label{ch1.sec2b.lemma2}
Let Theorem~\ref{ch1.sec1.thm1}(a) hold, and define the following Lyapunov functional in $D^{expl}(\infty)$,
\begin{eqnarray}
\tilde{V}(t)&=&V(t)+\beta\left[\int_{t_{0}}^{h_{2}}\int_{t_{0}}^{h_{1}}f_{T_{2}}(u)f_{T_{1}}(s)e^{-(\mu_{v}s+\mu u)}\int^{t}_{t-u}S(\theta)\frac{G(I(\theta-s))}{I(t)}d\theta dsdu\right.\nonumber\\
&&\left. +\int_{t_{0}}^{h_{2}}\int_{t_{0}}^{h_{1}}f_{T_{2}}(u)f_{T_{1}}(s)e^{-(\mu_{v}s+\mu u)}\int^{t}_{t-s}S(t)\frac{G(I(\theta))}{I(t)}d\theta\right],
\label{ch1.sec2b.lemma2.eq1}
\end{eqnarray}
where $V(t)$ is defined in (\ref{ch1.sec2a.thm1.proof.eq1}). It follows that
\begin{equation}\label{ch1.sec2b.lemma2.eq2}
\limsup_{t\rightarrow \infty}\frac{1}{t}\log{(I(t))}\leq \beta \frac{B}{\mu}E(e^{-(\mu_{v}T_{1}+\mu T_{2})})-(\mu+d+\alpha),\quad a.s.
\end{equation}
\end{lemma}
Proof:\\
The differential operator $dV$ applied to the Lyapunov functional $\tilde{V}(t)$
with respect to the system (\ref{ch1.sec0.eq8})-(\ref{ch1.sec0.eq11}) leads to the following
\begin{equation}\label{ch1.sec2b.lemma2.proof.eq1}
  d\tilde{V}(t)=f(S, I)dt  + \sigma_{\beta} \int_{t_{0}}^{h_{2}}\int^{h_{1}}_{t_{0}}f_{T_{2}}(u)f_{T_{1}}(s) e^{-(\mu_{v} s+\mu u)}S(t-u)\frac{G(I(t-s-u))}{I(t)}dsdudw_{\beta}(t),
\end{equation}
where,
\begin{eqnarray}
  f(S, I)&=&\beta\int_{t_{0}}^{h_{2}}f_{T_{2}}(u)\int^{h_{1}}_{t_{0}}f_{T_{1}}(s) e^{-(\mu_{v} s+\mu u)}S(t)\frac{G(I(t))}{I(t)}dsdu-(\mu+ d+ \alpha)\nonumber\\
  &&-\frac{1}{2}\sigma^{2}_{\beta}\left(\int_{t_{0}}^{h_{2}}f_{T_{2}}(u)\int^{h_{1}}_{t_{0}}f_{T_{1}}(s) e^{-(\mu_{v} s+\mu u)}S(t-u)\frac{G(I(t-s-u))}{I(t)}dsdu\right)^{2}.\label{ch1.sec2b.lemma2.proof.eq2}
\end{eqnarray}
Since $S(t), I(t)\in D^{expl}(\infty)$, and $G$ satisfies the conditions of Assumption~\ref{ch1.sec0.assum1}, it  follows easily that
\begin{equation}\label{ch1.sec2b.lemma2.proof.eq3}
f(S,I)\leq \beta \frac{B}{\mu}E(e^{-(\mu_{v}T_{1}+\mu T_{2})})-(\mu+d+\alpha).
\end{equation}
Now, integrating both sides of  (\ref{ch1.sec2b.lemma2.proof.eq1}) over the interval $[t_{0},t]$, it follows from (\ref{ch1.sec2a.thm1.proof.eq5})  and (\ref{ch1.sec2b.lemma2.eq1}) that
 \begin{eqnarray}
   \log{I(t)}&\leq&\tilde{V}(t)\nonumber\\
    &\leq&\tilde{V}(t_{0})+\left[\beta \frac{B}{\mu}E(e^{-(\mu_{v}T_{1}+\mu T_{2})})-(\mu+d+\alpha)\right](t-t_{0})+M_{2}(t),\label{ch1.sec2b.lemma2.proof.eq4}
 \end{eqnarray}
where $M_{2}(t)$ is defined in (\ref{ch1.sec2a.thm1.proof.eq5}).
 Diving both sides of (\ref{ch1.sec2b.lemma2.proof.eq4}) by $t$, and  taking the limit supremum  as $t\rightarrow\infty$, it is easy to see that (\ref{ch1.sec2b.lemma2.proof.eq4}) reduces to
 \begin{eqnarray}
   \limsup_{t\rightarrow\infty}\frac{1}{t}\log{I(t)}&\leq& \left[\beta \frac{B}{\mu}E(e^{-(\mu_{v}T_{1}+\mu T_{2})})-(\mu+d+\alpha)\right]
   +\limsup_{t\rightarrow\infty}\frac{1}{t}M_{2}(t).\label{ch1.sec2b.lemma2.proof.eq5}
 \end{eqnarray}
 But, from (\ref{ch1.sec2a.thm1.proof.eq5a}), applying Assumption~\ref{ch1.sec0.assum1} and $H\ddot{o}lder$ inequality to the quadratic variation of $M_{2}(t)$, it is easy to see that
 \begin{equation}\label{ch1.sec2b.lemma2.proof.eq5a}
  <M_{2}(t), M_{2}(t)>\leq \sigma^{2}_{\beta}\int^{t}_{t_{0}} \int_{t_{0}}^{h_{2}}\int^{h_{1}}_{t_{0}}f_{T_{2}}(u)f_{T_{1}}(s) e^{-2(\mu_{v} s+\mu u)}S^{2}(\xi-u)\frac{(I(\xi-s-u))^{2}}{I^{2}(\xi)}dsdud\xi.
\end{equation}
Furthermore, in $D^{expl}(\infty)$,
 \begin{equation}\label{ch1.sec2b.lemma2.proof.eq6}
 \frac{\left(\frac{B}{\mu+d}\right)^{4}}{\left(\frac{B}{\mu}\right)^{2}}\leq \frac{S^{2}(\xi-u)I^{2}(\xi-s-u)}{I^{2}(\xi)}\leq \frac{\left(\frac{B}{\mu}\right)^{4}}{\left(\frac{B}{\mu+d}\right)^{2}}, \forall \xi\in [t_{0},t],s\in[t_{0},T_{1}], u\in[t_{0}, T_{2}].
 \end{equation}
 Thus, from (\ref{ch1.sec2b.lemma2.proof.eq5a}), the quadratic variation of $M_{2}(t)$ satisfies
 \begin{equation}\label{ch1.sec2b.lemma2.proof.eq7}
 \limsup_{t\rightarrow \infty}\frac{1}{t}<M_{2}(t), M_{2}(t)>\leq \sigma^{2}_{\beta}\frac{\left(\frac{B}{\mu}\right)^{4}}{\left(\frac{B}{\mu+d}\right)^{2}}E(e^{-2(\mu_{v}T_{1}+\mu T_{2})})<\infty.
\end{equation}
Therefore, it is easy to see by the strong law of large numbers for local martingales (see, e.g. \cite{mao}) that
\begin{equation}\label{ch1.sec2b.lemma2.proof.eq8}
  \limsup_{t\rightarrow \infty}{\frac{1}{t}M_{2}(t)}=0,\quad a.s.
\end{equation}
And the result (\ref{ch1.sec2b.lemma2.eq2}) follows immediately from (\ref{ch1.sec2b.lemma2.proof.eq8}) and (\ref{ch1.sec2b.lemma2.proof.eq5}).

The conditions for extinction of the infectious population over time can be expressed in terms of two important parameters for the disease dynamics namely - (1) the basic reproduction number $R^{*}_{0}$ in (\ref{ch1.sec2.lemma2a.corrolary1.eq4}), and (2) the expected survival probability rate of the parasites $E(e^{-(\mu_{v}T_{1}+\mu T_{2})})$, defined in [Theorem~5.1, Wanduku\cite{wanduku-biomath}].
\begin{thm}\label{ch1.sec2b.thm1}
Suppose the conditions for Lemma~\ref{ch1.sec2b.lemma2} are satisfied, and let the basic reproduction number $R^{*}_{0}$ be defined as in (\ref{ch1.sec2.lemma2a.corrolary1.eq4}). In addition, let one of the following conditions hold
\item[1.] $R^{*}_{0}\geq 1$ and $E(e^{-(\mu_{v}T_{1}+\mu T_{2})})<\frac{1}{R^{*}_{0}}$, or
\item[2.]$R^{*}_{0}<1$.\\
Then
\begin{equation}\label{ch1.sec2b.thm1.eq1}
\limsup_{t\rightarrow \infty}\frac{1}{t}\log{(I(t))}<-\lambda, \quad a.s.
\end{equation}
where  $\lambda>0$ is some positive constant. In other words, $I(t)$ converges to zero exponentially, almost surely.
\end{thm}
Proof:\\
Suppose Theorem~\ref{ch1.sec2b.thm1}~[1.] holds, then from (\ref{ch1.sec2b.lemma2.eq2}),
\begin{equation}\label{ch1.sec2b.thm1.eq1.proof.eq1}
\limsup_{t\rightarrow \infty}\frac{1}{t}\log{(I(t))}< \beta\frac{B}{\mu}\left(E(e^{-(\mu_{v}T_{1}+\mu T_{2})})- \frac{1}{R^{*}_{0}} \right)\equiv -\lambda,
\end{equation}
where the positive constant $\lambda>0$ is taken to be  as follows
\begin{equation}\label{ch1.sec2b.thm1.eq1.proof.eq1.eq1}
\lambda\equiv(\mu+d+\alpha)-\beta \frac{B}{\mu}E(e^{-(\mu_{v}T_{1}+\mu T_{2})})=\beta\frac{B}{\mu}\left( \frac{1}{R^{*}_{0}}-E(e^{-(\mu_{v}T_{1}+\mu T_{2})}) \right)>0.
\end{equation}
 Also, suppose Theorem~\ref{ch1.sec2b.thm1}~[2.] holds, then from (\ref{ch1.sec2b.lemma2.eq2}),
\begin{eqnarray}
\limsup_{t\rightarrow \infty}\frac{1}{t}\log{(I(t))}&\leq& \beta \frac{B}{\mu}E(e^{-(\mu_{v}T_{1}+\mu T_{2})})-(\mu+d+\alpha)\nonumber\\
&<& \beta \frac{B}{\mu}-(\mu+d+\alpha)= -(1-R^{*}_{0})(\mu+d+\alpha)\equiv -\lambda,\label{ch1.sec2b.thm1.eq1.proof.eq2}
\end{eqnarray}
where the positive constant $\lambda>0$ is taken to be  as follows
\begin{equation}\label{ch1.sec2b.thm1.eq1.proof.eq2.eq1}
\lambda\equiv(1-R^{*}_{0})(\mu+d+\alpha)>0.
\end{equation}
\begin{rem}
Theorem~\ref{ch1.sec2b.thm1}, Theorem~\ref{ch1.sec1.thm1}[a] and Lemma~\ref{ch1.sec2b.lemma1} signify that when the intensity  of the noise from the disease transmission rate $\sigma_{\beta}$  is positive, and the intensities of the noises from the natural death rates  satisfy $\sigma_{i}=0, i\in \{S, E, I, R\}$,   then all sample paths of the solution process $\{(S(t), I(t)),t\geq t_{0}\}$ of the decoupled system (\ref{ch1.sec0.eq9}) and (\ref{ch1.sec0.eq11}) that start  in $D^{expl}(\infty)\subset D(\infty)$ continue to oscillate in $D^{expl}(\infty)$. Moreover, the sample paths of the infectious state $I(t), t\geq t_{0}$ of the solution process $\{(S(t), I(t)),t\geq t_{0}\}$ ultimately turn to zero exponentially, almost surely, whenever either the expected survival probability rate of the malaria parasites satisfy  $E(e^{-(\mu_{v}T_{1}+\mu T_{2})})<\frac{1}{R^{*}_{0}}$, or whenever the basic production number of the disease satisfy $R^{*}_{0}<1$.  Furthermore, the sample Lyapunov exponent from (\ref{ch1.sec2b.thm1.eq1}) is estimated by the  term $\lambda$, defined in (\ref{ch1.sec2b.thm1.eq1.proof.eq1.eq1}) and (\ref{ch1.sec2b.thm1.eq1.proof.eq2.eq1}).

  It follows from (\ref{ch1.sec2b.thm1.eq1}) that when either of the conditions in Theorem~\ref{ch1.sec2b.thm1}[1.-2.]  hold, then the infectious population $I(t)$ dies out exponentially, almost surely, whenever  $\lambda$ in (\ref{ch1.sec2b.thm1.eq1.proof.eq1.eq1}) and (\ref{ch1.sec2b.thm1.eq1.proof.eq2.eq1}) is positive, that is, $\lambda>0$. In addition, the rate of the exponential decay of each sample path of the infectious population $I(t)$ in each scenario of Theorem~\ref{ch1.sec2b.thm1}[1.-2.] is given by the  estimate $\lambda>0$ of the sample Lyapunov exponent in (\ref{ch1.sec2b.thm1.eq1.proof.eq1.eq1}) and (\ref{ch1.sec2b.thm1.eq1.proof.eq2.eq1}).

  The conditions in  Theorem~\ref{ch1.sec2b.thm1}[1.-2.] can also be interpreted as follows. Recall [\cite{wanduku-biomath}, Remark~4.2], the basic reproduction number   $R^{*}_{0}$ in (\ref{ch1.sec2.lemma2a.corrolary1.eq4}) (similarly in (\ref{ch1.sec2.theorem1.corollary1.eq3})) represents the expected number of secondary malaria cases that result from one infective placed in the steady state disease free population $S^{*}_{0}=\frac{B}{\mu}$. Thus, $\frac{1}{R^{*}_{0}}=\frac{(\mu+d+\alpha)}{\beta S^{*}_{0}}$, for $R^{*}_{0}\geq 1$, represents the probability rate of  infectious persons in the secondary infectious population $\beta S^{*}_{0}$ leaving the infectious state either through natural death $\mu$, diseases related death $d$, or recovery and acquiring natural immunity at the rate $\alpha$. Thus, $\frac{1}{R^{*}_{0}}$ is the effective probability rate of surviving infectiousness until recovery with acquisition of natural immunity. Moreover, $\frac{1}{R^{*}_{0}}$ is a probability measure provided $R^{*}_{0}\geq 1$.

  In addition, recall [\cite{wanduku-biomath}, Theorem~5.1\&5.2] asserts that when $R^{*}_{0}\geq 1$, and the expected survival probability $E(e^{-(\mu_{v}T_{1}+\mu T_{2})})$ is significantly large, then the outbreak of  malaria establishes a malaria endemic steady state population $E_{1}$. The conditions for extinction of disease in  Theorem~\ref{ch1.sec2b.thm1}[1.], that is $R^{*}_{0}\geq 1$ and $E(e^{-(\mu_{v}T_{1}+\mu T_{2})})<\frac{1}{R^{*}_{0}}$ suggest that in the event where $R^{*}_{0}\geq 1$, and the disease is aggressive, and likely to establish an endemic steady state population, if the expected survival probability rate $E(e^{-(\mu_{v}T_{1}+\mu T_{2})})$ of the malaria parasites over their complete life cycle of length $T_{1}+T_{2}$, is less than $\frac{1}{R^{*}_{0}}$- the effective probability rate of surviving infectiousness until recovery with natural immunity, then the malaria epidemic fails to establish an endemic steady state, and as a result, the disease ultimately dies out at an exponential rate $\lambda$ in (\ref{ch1.sec2b.thm1.eq1.proof.eq1.eq1}).
  This result suggests that, malaria control policies should embark on vector control strategies such as genetic modification techniques in order to reduce the chances of survival of the malaria parasites inside the mosquitos, and in the human beings.

   In the event where  $R^{*}_{0}< 1$ in  Theorem~\ref{ch1.sec2b.thm1}[2.], extinction of disease occurs exponentially over sufficiently long time, regardless of the survival of the parasites. Moreover, the rate of extinction is  $\lambda$ in (\ref{ch1.sec2b.thm1.eq1.proof.eq2.eq1}).
 Also observe that the conditions in Theorem~\ref{ch1.sec2b.thm1}[1.-2.] for extinction of the infectious population $I(t)$ in the case of noise originating exclusively from the disease transmission rate $\beta$ has no bearings on the intensity of the  noise from the disease transmission rate $\sigma_{\beta}$.
\end{rem}
As it can be observed  from  several simulation studies involving white noise processes, in many occasions, the extinction of the infectious population over time coincides with extinction of the susceptible population, if the intensity of the noise in the epidemic dynamic system is high. And this suggests that the extinction of the disease from the population does not always imply the survival of the disease free population over time.

 The following result describes the average behavior of the trajectories of the susceptible population over sufficiently long time in the phase plane of the solution process $\{(S(t), I(t)), t\geq t_{0}\}$ of the decoupled system (\ref{ch1.sec0.eq9}) and (\ref{ch1.sec0.eq11}), and also states conditions for the asymptotic stability in the mean of the trajectories (see Definition~\ref{ch1.sec2a.defn1}(2)), in the event where the conditions of Theorem~\ref{ch1.sec2b.thm1} are satisfied.
\begin{thm}\label{ch1.sec2b.thm2}
Suppose any of the conditions in the hypothesis of Theorem~\ref{ch1.sec2b.thm1}[1.-2.] are satisfied. It follows that in $D^{expl}(\infty)$, the paths of the susceptible population in the solution process $\{(S(t), I(t)), t\geq t_{0}\}$ of the decoupled system (\ref{ch1.sec0.eq9}) and (\ref{ch1.sec0.eq11}) satisfy
\begin{equation}\label{ch1.sec2b.thm2.eq1}
\lim_{t\rightarrow \infty}\frac{1}{t}\int_{t_{0}}^{t}S(\xi)d\xi=\frac{B}{\mu},\quad a.s.
\end{equation}
That is, the susceptible population is strongly persistent over long-time in the mean, and almost sure asymptotically stable on average. Moreover, the average value of the susceptible population over sufficiently long time is the disease-free equilibrium $S^{*}_{0}=\frac{B}{\mu}$.
\end{thm}
Proof:\\
Suppose either of the conditions in Theorem~\ref{ch1.sec2b.thm1}[1.-2.] hold, then if we further let
\begin{equation}\label{ch1.sec2b.thm2.proof.eq1}
  \Omega_{1}=\{w\in \Omega: \limsup_{t\rightarrow \infty} I(w,t)=0\},
\end{equation}
then it follows clearly from Theorem~\ref{ch1.sec2b.thm1} that $P(\Omega_{1})=1$. That is, for every $\epsilon>0$, there is a positive constant $K_{1}(w,\epsilon)\equiv K_{1}>0$, such that
\begin{equation}\label{ch1.sec2b.thm2.proof.eq2}
  I(w,t)<\epsilon,\quad a.s.\quad \forall w\in \Omega_{1},\quad \textrm{whenever $t>K_{1}$}.
\end{equation}
It follows from (\ref{ch1.sec2b.thm2.proof.eq2}) that
\begin{equation}\label{ch1.sec2b.thm2.proof.eq3}
  I(w,t-s)<\epsilon,\quad a.s.\quad \forall w\in \Omega_{1},\quad \textrm{whenever $t>K_{1}+h_{1},\forall s\in [t_{0}, h_{1}]$}.
\end{equation}
In $D^{expl}(\infty)$, define
\begin{equation}\label{ch1.sec2b.thm2.proof.eq4}
V_{1}(t)=S(t)+\alpha \int_{t_{0}}^{\infty} f_{T_{3}}(r)e^{\mu r}\int_{t-r}^{t}I(\theta)d\theta dr.
\end{equation}
The differential operator $dV_{1}$ applied to the Lyapunov functional $V_{1}(t)$ in (\ref{ch1.sec2b.thm2.proof.eq4}) leads to the following
\begin{equation}\label{ch1.sec2b.thm2.proof.eq5}
  dV_{1}(t)=\left[g(S, I)-\mu S(t)\right]dt-\sigma_{\beta} S(t)\int^{h_{1}}_{t_{0}}f_{T_{1}}(s) e^{-\mu_{v} s}G(I(t-s))dsdw_{\beta}(t),
\end{equation}
where
\begin{equation}\label{ch1.sec2b.thm2.proof.eq6}
  g(S,I)=B-\beta S(t)\int^{h_{1}}_{t_{0}}f_{T_{1}}(s) e^{-\mu_{v} s}G(I(t-s))ds+\alpha E(e^{-\mu T_{3}}) I(t).
\end{equation}
Estimating the right-hand-side of (\ref{ch1.sec2b.thm2.proof.eq5}) in $D^{expl}(\infty)$, and integrating over $[t_{0},t]$, it follows from (\ref{ch1.sec2b.thm2.proof.eq2})-(\ref{ch1.sec2b.thm2.proof.eq3}) that
\begin{eqnarray}
  V_{1}(t)&\leq& V_{1}(t_{0})+B(t-t_{0})+\int_{t_{0}}^{K_{1}}\alpha I(\xi)d\xi +\int_{K_{1}}^{t}\alpha I(\xi)d\xi -\mu\int_{t_{0}}^{t} S(\xi)d\xi-M_{3}(t),\nonumber\\
  &\leq& V_{1}(t_{0})+B(t-t_{0})+\alpha \frac{B}{\mu}(K_{1}-t_{0})  +\alpha (t-K_{1})\epsilon -\mu\int_{t_{0}}^{t} S(\xi)d\xi-M_{3}(t),\label{ch1.sec2b.thm2.proof.eq7}
\end{eqnarray}
where
\begin{equation}\label{ch1.sec2b.thm2.proof.eq8}
  M_{3}(t)=\sigma_{\beta}\int_{t_{0}}^{t} S(\xi)\int^{h_{1}}_{t_{0}}f_{T_{1}}(s) e^{-\mu_{v} s}G(I(\xi-s))dsdw_{\beta}(\xi).
\end{equation}
Observe that similarly to (\ref{ch1.sec2b.lemma2.proof.eq5a})-(\ref{ch1.sec2b.lemma2.proof.eq8}), it is easy to see by the strong law of large numbers for local martingales (see, e.g. \cite{mao}) that
\begin{equation}\label{ch1.sec2b.thm2.proof.eq9}
  \lim_{t\rightarrow \infty}{\frac{1}{t}M_{3}(t)}=0,\quad a.s.
\end{equation}
Thus, dividing both sides of (\ref{ch1.sec2b.thm2.proof.eq7}) by $t$ and taking the limit supremum as $t\rightarrow \infty$, it follows that
\begin{equation}\label{ch1.sec2b.thm2.proof.eq10}
\limsup_{t\rightarrow \infty}\frac{1}{t}\int_{t_{0}}^{t}S(\xi)d\xi\leq \frac{B}{\mu}+ \frac{\alpha}{\mu}\epsilon,\quad a.s.
\end{equation}
On the other hand, estimating $g(S,I)$ in (\ref{ch1.sec2b.thm2.proof.eq6}) from below and using the conditions of Assumption~\ref{ch1.sec0.assum1} and (\ref{ch1.sec2b.thm2.proof.eq3}), it is easy to see that  in $D^{expl}(\infty)$,
\begin{eqnarray}
  g(S,I) &\geq & B-\beta S(t)\int^{h_{1}}_{t_{0}}f_{T_{1}}(s) e^{-\mu_{v} s}(I(t-s))ds\nonumber \\
   &\geq& B-\beta \frac{B}{\mu}E(e^{-\mu_{v}T_{1}})\epsilon,\forall w\in\Omega_{1}\quad and\quad t>K_{1}+h_{1},\nonumber\\
   &\geq&B-\beta \frac{B}{\mu}\epsilon.\label{ch1.sec2b.thm2.proof.eq11}
\end{eqnarray}
Moreover, for $t\in[t_{0}, K_{1}+h_{1}]$, then
\begin{equation}\label{ch1.sec2b.thm2.proof.eq11.eq1}
  g(S,I)\geq B-\beta \left(\frac{B}{\mu}\right)^{2}.
\end{equation}
Therefore, applying (\ref{ch1.sec2b.thm2.proof.eq11})-(\ref{ch1.sec2b.thm2.proof.eq11.eq1}) into (\ref{ch1.sec2b.thm2.proof.eq5}), then integrating both sides of (\ref{ch1.sec2b.thm2.proof.eq5}) over $[t_{0},t]$, and diving the result by $t$,  it is easy to see from (\ref{ch1.sec2b.thm2.proof.eq5}) that
\begin{equation}\label{ch1.sec2b.thm2.proof.eq12}
  \frac{1}{t}V_{1}(t)\geq  \frac{1}{t}V_{1}(t_{0})+B(1-\frac{t_{0}}{t}) -\frac{1}{t}\beta\left( \frac{B}{\mu}\right)^{2}(K_{1}+h_{1}-t_{0})-\beta\frac{B}{\mu}\epsilon[1-\frac{K_{1}+h_{1}}{t}] -\frac{1}{t}\mu\int_{t_{0}}^{t}S(\xi)d\xi-\frac{1}{t}M_{3}(t).
\end{equation}
Observe that in  $D^{expl}(\infty)$, $\lim_{t\rightarrow \infty}\frac{1}{t}V_{1}(t)=0$, a.s., and $\lim_{t\rightarrow \infty}\frac{1}{t}V_{1}(t_{0})=0$. Moreover, from (\ref{ch1.sec2b.thm2.proof.eq9}), $\lim_{t\rightarrow \infty}{\frac{1}{t}M_{3}(t)}=0,\quad a.s$. Therefore, rearranging (\ref{ch1.sec2b.thm2.proof.eq12}), and taking the limit infinimum of both sides as $t\rightarrow \infty$, it is easy to see that
\begin{equation}\label{ch1.sec2b.thm2.proof.eq13}
 \liminf_{t\rightarrow \infty} \frac{1}{t}\int_{t_{0}}^{t}S(\xi)d\xi\geq   \frac{B}{\mu}-\frac{1}{\mu}\beta \frac{B}{\mu}\epsilon,\quad a.s.
\end{equation}
It follows from (\ref{ch1.sec2b.thm2.proof.eq10}) and (\ref{ch1.sec2b.thm2.proof.eq13}) that
\begin{equation}\label{ch1.sec2b.thm2.proof.eq14}
\frac{B}{\mu}-\frac{1}{\mu}\beta \frac{B}{\mu}\epsilon\leq \liminf_{t\rightarrow \infty} \frac{1}{t}\int_{t_{0}}^{t}S(\xi)d\xi\leq \lim_{t\rightarrow \infty} \frac{1}{t}\int_{t_{0}}^{t}S(\xi)d\xi\leq \limsup_{t\rightarrow \infty}\frac{1}{t}\int_{t_{0}}^{t}S(\xi)d\xi\leq \frac{B}{\mu}+ \frac{\alpha}{\mu}\epsilon,\quad a.s.
\end{equation}
Hence, for $\epsilon$ arbitrarily small,  the result in (\ref{ch1.sec2b.thm2.eq1}) follows immediately from (\ref{ch1.sec2b.thm2.proof.eq14}).
\begin{rem}\label{ch1.sec2b.thm2.rem1}
Theorem~\ref{ch1.sec2b.thm2}, Theorem~\ref{ch1.sec2b.thm1}, Theorem~\ref{ch1.sec1.thm1}[a] and Lemma~\ref{ch1.sec2b.lemma1} signify that when the intensity  of the noise from the disease transmission rate $\sigma_{\beta}$  is positive, and the intensities of the noises from the natural death rates  satisfy $\sigma_{i}=0, i\in \{S, E, I, R\}$,   then all sample paths of the solution process $\{(S(t), I(t)),t\geq t_{0}\}$ of the decoupled system (\ref{ch1.sec0.eq9}) and (\ref{ch1.sec0.eq11}) that start  in $D^{expl}(\infty)\subset D(\infty)$ continue to oscillate in $D^{expl}(\infty)$. Moreover, the sample paths of the infectious state $I(t), t\geq t_{0}$ of the solution process $\{(S(t), I(t)),t\geq t_{0}\}$ ultimately turn to zero exponentially, almost surely, whenever either the expected survival probability rate the malaria parasite satisfy  $E(e^{-(\mu_{v}T_{1}+\mu T_{2})})<\frac{1}{R^{*}_{0}}$, for $R^{*}_{0}\geq 1$, or whenever the basic production number satisfy $R^{*}_{0}<1$.  Furthermore, the rate of the exponential decrease of the infectious population from (\ref{ch1.sec2b.thm1.eq1}) is estimated by the  term $\lambda$, defined in (\ref{ch1.sec2b.thm1.eq1.proof.eq1.eq1}) and (\ref{ch1.sec2b.thm1.eq1.proof.eq2.eq1}).

In addition, Theorem~\ref{ch1.sec2b.thm2} asserts that when either the expected survival probability rate the malaria parasites satisfy  $E(e^{-(\mu_{v}T_{1}+\mu T_{2})})<\frac{1}{R^{*}_{0}}$, for $R^{*}_{0}\geq 1$, or whenever the basic production number satisfy $R^{*}_{0}<1$, the susceptible population remains persistent in the mean over sufficiently large time, moreover, every sample path of the susceptible population $S(t)$ that starts in $D^{expl}(\infty)$ continues to oscillate in $D^{expl}(\infty)$, and  on average all sample paths converge to the disease free steady state population $S^{*}_{0}=\frac{B}{\mu}$.

In other words, over sufficiently long time, the population that remains will be all susceptible malaria-free people, and the population size will be equal to the disease free steady state population $S^{*}_{0}=\frac{B}{\mu}$ of the system (\ref{ch1.sec0.eq8})-(\ref{ch1.sec0.eq11}). See $S^{*}_{0}=\frac{B}{\mu}$ in Wanduku\cite{wanduku-biomath}.
\end{rem}
Based on the observations from Theorem~\ref{ch1.sec2b.thm2}, the following proposition is made, and the proof given elsewhere.
\begin{prop}\label{ch1.sec2b.thm2.prop1}
Suppose the assumptions of Theorem~\ref{ch1.sec2b.thm2} are satisfied, then the disease-free equilibrium $E_{0}=(S^{*}_{0},0)=(\frac{B}{\mu},0)$ of the decoupled system (\ref{ch1.sec0.eq9}) and (\ref{ch1.sec0.eq11}) is stochastically asymptotically stable in the large. In other words, if either the basic reproduction number  $R^{*}_{0}$, or the expected survival probability of the malaria plasmodium $E(e^{-(\mu_{v}T_{1}+\mu T_{2})})$,  satisfy $R^{*}_{0}<1$, or $E(e^{-(\mu_{v}T_{1}+\mu T_{2})})<\frac{1}{R^{*}_{0}}$, whenever $R^{*}_{0}\geq 1$, respectively, then every trajectory for the solution process $\{(S(t), I(t)), t\geq t_{0}\}$ of the decoupled system (\ref{ch1.sec0.eq9}) and (\ref{ch1.sec0.eq11}) that starts near $E_{0}$, remains near $E_{0}$, and converges asymptotically to $E_{0}$, almost surely.
\end{prop}
 \section{Example}\label{ch1.sec4}
The examples exhibited in this section are used to facilitate understanding about the influence of the intensity or "strength" of the noise in the system on the extinction of the disease in the population over time. This objective is achieved in a simplistic manner by examining the behavior of the sample paths of  the different states ($S, E, I, R$) of the stochastic system (\ref{ch1.sec0.eq8})-(\ref{ch1.sec0.eq11}) over sufficiently long time.
%
 \subsection{Example 1: The joint effect of the intensity of white noise from disease transmission and natural deathrates}\label{ch1.sec4.subsec1}
The following convenient list of parameter values in Table~\ref{ch1.sec4.table2} are used to generate and examine the paths of the different states of the stochastic system (\ref{ch1.sec0.eq8})-(\ref{ch1.sec0.eq11}), whenever the conditions of Theorem~\ref{ch1.sec2a.thm1} are satisfied.
 \begin{table}[h]
  \centering
  \caption{A list of specific values chosen for the system parameters for the examples in subsection~\ref{ch1.sec4.subsec1}}\label{ch1.sec4.table2}
  \begin{tabular}{l l l}
  Disease transmission rate&$\beta$& 0.0006277\\\hline
  Constant Birth rate&$B$&$ \frac{22.39}{1000}$\\\hline
  Recovery rate& $\alpha$& 0.55067\\\hline
  Disease death rate& $d$& 0.011838\\\hline
  Natural death rate& $\mu$& $0.6$\\\hline
  Incubation delay time in vector& $T_{1}$& 2 units \\\hline
  Incubation delay time in host& $T_{2}$& 1 unit \\\hline
  Immunity delay time& $T_{3}$& 4 units\\\hline
  \end{tabular}
\end{table}
The Euler-Maruyama stochastic approximation scheme\footnote{A seed is set on the random number generator to reproduce  the same sequence of random numbers for the Brownian motion in order to generate reliable graphs for the trajectories of the system under different intensity values for the white noise processes, so that comparison can be made to identify differences that reflect the effect of intensity values.} is used to generate trajectories for the different states $S(t), E(t), I(t), R(t)$ over the time interval $[0,T]$, where $T=\max(T_{1}+T_{2}, T_{3})=4$. The special nonlinear incidence  functions $G(I)=\frac{aI}{1+I}, a=0.05$ in \cite{gumel} is utilized to generate the numeric results. Furthermore, the following initial conditions are used
\begin{equation}\label{ch1.sec4.eq1}
\left\{
\begin{array}{l l}
S(t)= 10,\\
E(t)= 5,\\
I(t)= 6,\\
R(t)= 2,
\end{array}
\right.
\forall t\in [-T,0], T=\max(T_{1}+T_{2}, T_{3})=4.
\end{equation}
\begin{figure}[H]
\begin{center}
\includegraphics[height=6cm]{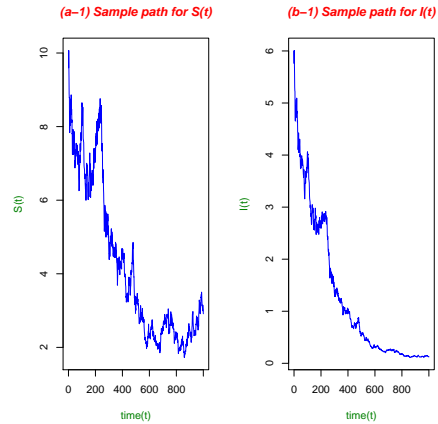}
\caption{(a-1), and (b-1),  show the trajectories of the  states $(S,I)$, respectively, over sufficiently long time $t=1000$, whenever the intensity of the incidence of malaria is $a=0.05$, and the intensities of the white noise processes take the values $\sigma_{i}=0.05,\forall i\in\{S, E, I, R\}$, and $\sigma_{\beta}=2.5$.
Also, note that all  negative values have no meaningful interpretation, except a significance of extinction of the population if the negative values occur over sufficiently long time. Moreover, the basic reproduction number in (\ref{ch1.sec2.lemma2a.corrolary1.eq4}) in this case is $R^{*}_{0}= 2.014926e-05<1$, the estimate Lyapunov exponent or rate of extinction of the disease in (\ref{ch1.sec2a.rem1.eq2}) is $Q=1.287508>0$
}\label{ch1.sec4.subsec1.fig1}
\end{center}
\end{figure}
Figure~\ref{ch1.sec4.subsec1.fig1} can be used to verify the results about the extinction of the infectious population in Theorem~\ref{ch1.sec2a.thm1}. Indeed, it can be observed that for the given parameter values in Table~\ref{ch1.sec4.table2}, and the initial conditions for the system (\ref{ch1.sec0.eq8})-(\ref{ch1.sec0.eq11}) in (\ref{ch1.sec4.eq1}), and the intensities of the white noise processes in the system $\sigma_{i}=0.05,\forall i\in\{S, E, I, R\}$, and $\sigma_{\beta}=2.5$, it follows from (\ref{ch1.sec2a.rem1.eq1}) that the estimate of the  Lyapunov exponent or the rate of extinction of the malaria population $I(t)$ is $Q=1.287508>0$. That is,
\begin{equation}\label{ch1.sec4.subsec1.eq1}
  \limsup_{t\rightarrow \infty}{\frac{1}{t}\log{(I(t))}}\leq -Q = -1.287508 \quad a.s.
\end{equation}
The Figure~\ref{ch1.sec4.subsec1.fig1}(b-1) confirms that over sufficiently large time, when $Q>0$, then the infectious population becomes extinct. Furthermore, note that the basic reproduction number in (\ref{ch1.sec2.lemma2a.corrolary1.eq4}) in this scenario is $R^{*}_{0}= 2.014926e-05<1$, which signifies that the disease is getting eradicated from the population over time, leading to the small rise in the susceptible population seen in Figure~\ref{ch1.sec4.subsec1.fig1}(a-1) over sufficiently long time. The general decrease in the susceptible population $S(t)$ in Figure~\ref{ch1.sec4.subsec1.fig1}(a-1) over time is accounted for  by the presence of noise in the natural death rate with a significant intensity $\sigma_{S}=0.05$.
 \subsection{Example 2: The effect of the intensity of white noise from disease transmission rate and stochastic stability}\label{ch1.sec4.subsec2}
The list of parameter values in Table~\ref{ch1.sec4.table2} are also used to examine the paths of the different states of the stochastic system (\ref{ch1.sec0.eq8})-(\ref{ch1.sec0.eq11}), whenever the conditions of Theorem~\ref{ch1.sec2b.thm1} and Theorem~\ref{ch1.sec2b.thm2} are satisfied.
\begin{figure}[H]
\begin{center}
\includegraphics[height=6cm]{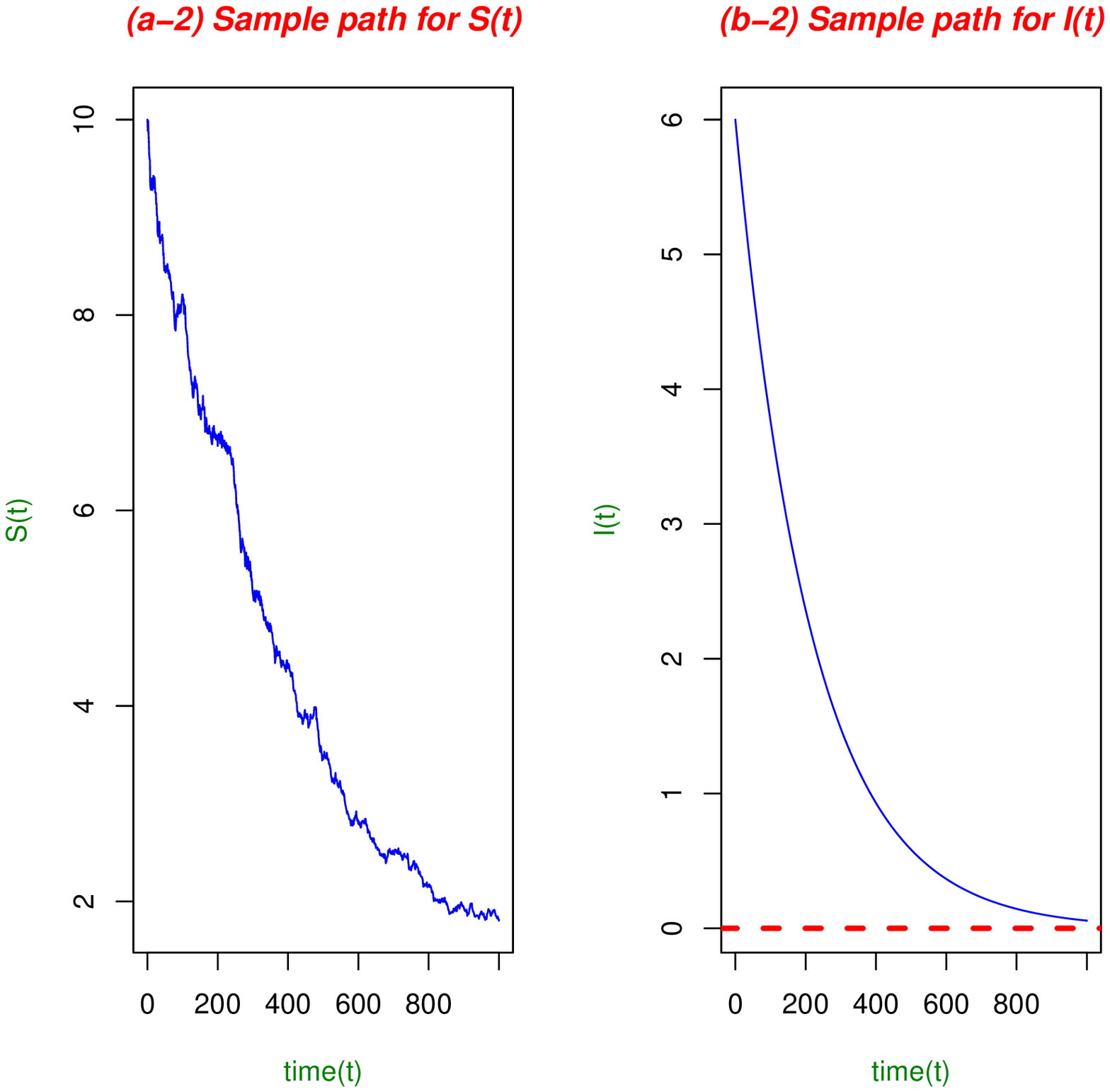}
\caption{(a-2), and (b-2),  show the trajectories of the  states $(S,I)$, respectively, over sufficiently long time $t=1000$, whenever the intensity of the incidence of malaria is $a=0.05$, and the intensities of the white noise processes take the values $\sigma_{i}=0,\forall i\in\{S, E, I, R\}$, and $\sigma_{\beta}=10$.
Also, note that all  negative values have no meaningful interpretation, except a significance of extinction of the population if the negative values occur over sufficiently long time. Moreover, the basic reproduction number in (\ref{ch1.sec2.lemma2a.corrolary1.eq4}) in this case is $R^{*}_{0}= 2.014926e-05<1$, the estimate of the Lyapunov exponent or rate of extinction of the disease in (\ref{ch1.sec2b.thm1.eq1.proof.eq2.eq1}) is $\lambda= 1.162485>0$. The broken line in (b-2) signify the origin, while the broken line in (a-2) signify the disease-free-equilibrium $S^{*}_{0}=\frac{B}{\mu}=0.03731667$.
}\label{ch1.sec4.subsec1.fig2}
\end{center}
\end{figure}
   \begin{figure}[H]
\begin{center}
\includegraphics[height=6cm]{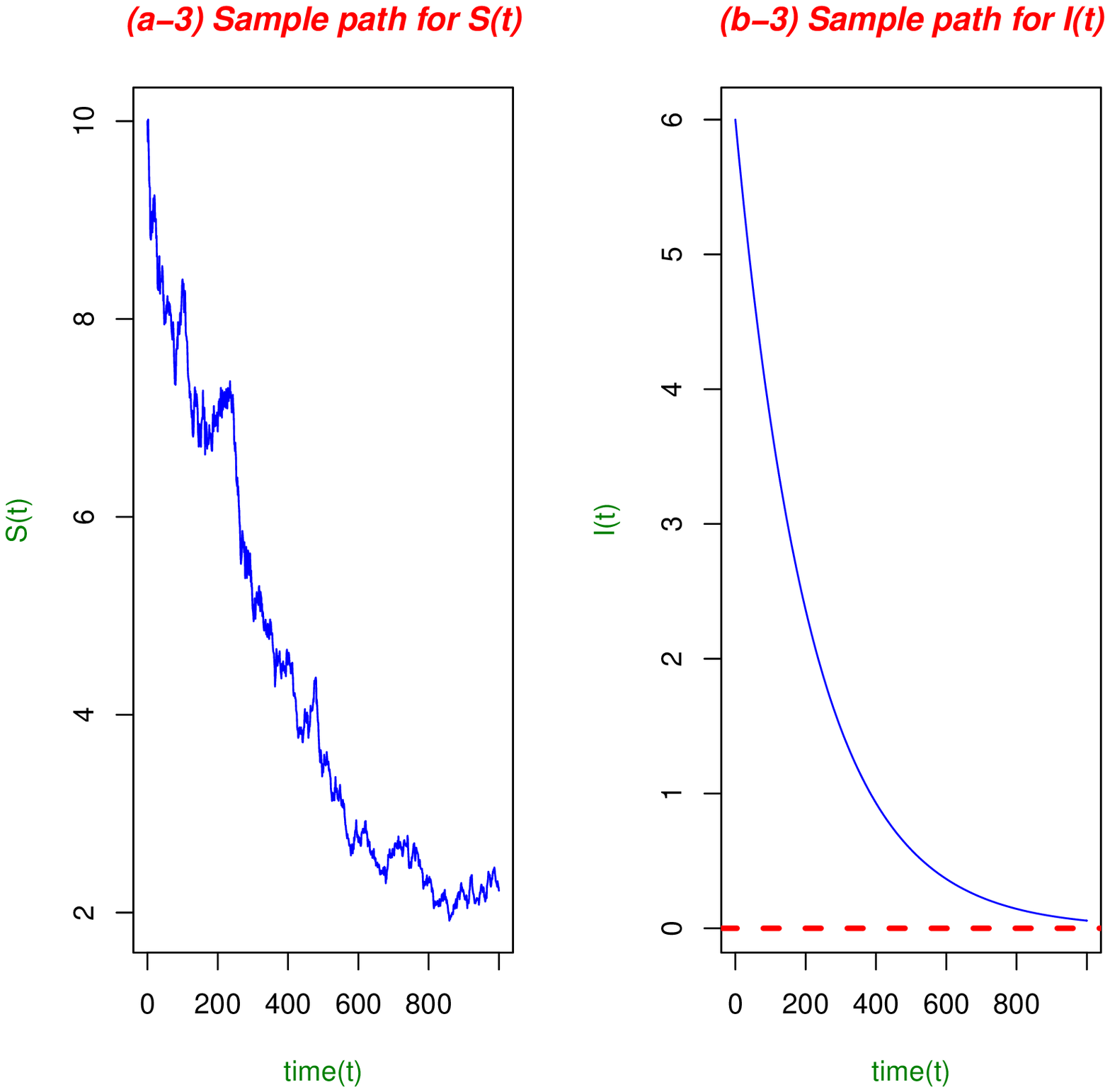}
\caption{(a-3), and (b-3),  show the trajectories of the  states $(S,I)$, respectively, over sufficiently long time $t=1000$, whenever the intensity of the incidence of malaria is $a=0.05$, and the intensities of the white noise processes take the values $\sigma_{i}=0,\forall i\in\{S, E, I, R\}$, and $\sigma_{\beta}=20$.
Also, note that all  negative values have no meaningful interpretation, except a significance of extinction of the population if the negative values occur over sufficiently long time. Moreover, the basic reproduction number in (\ref{ch1.sec2.lemma2a.corrolary1.eq4}) in this case is $R^{*}_{0}= 2.014926e-05<1$, the estimate of the Lyapunov exponent or rate of extinction of the disease in (\ref{ch1.sec2b.thm1.eq1.proof.eq2.eq1}) is $\lambda= 1.162485>0$. The broken line in (b-3) signify the origin, while the broken line in (a-3) signify the disease-free-equilibrium $S^{*}_{0}=\frac{B}{\mu}=0.03731667$.
}\label{ch1.sec4.subsec1.fig3}
\end{center}
\end{figure}

  Figure~\ref{ch1.sec4.subsec1.fig2} is used to verify the results about the extinction of the infectious population over time in Theorem~\ref{ch1.sec2b.thm1}, and the long-term behavior of the  susceptible population $S(t)$ in Theorem~\ref{ch1.sec2b.thm2}. Indeed, it can be observed that for the given parameter values in Table~\ref{ch1.sec4.table2}, and the initial conditions for the system (\ref{ch1.sec0.eq8})-(\ref{ch1.sec0.eq11}) in (\ref{ch1.sec4.eq1}), and the intensities of the white noise processes in the system $\sigma_{i}=0,\forall i\in\{S, E, I, R\}$, and $\sigma_{\beta}=10$, it follows that the basic reproduction number in (\ref{ch1.sec2.lemma2a.corrolary1.eq4}) in this scenario is $R^{*}_{0}= 2.014926e-05<1$. Therefore, the condition of Theorem~\ref{ch1.sec2b.thm1}(a.) is satisfied, and from (\ref{ch1.sec2b.thm1.eq1.proof.eq2.eq1}), the estimate of the  Lyapunov exponent or the rate of extinction of the malaria population $I(t)$ is $\lambda= 1.162485>0$. That is,
\begin{equation}\label{ch1.sec4.subsec1.eq1}
  \limsup_{t\rightarrow \infty}{\frac{1}{t}\log{(I(t))}}\leq -\lambda = -1.162485 \quad a.s.
\end{equation}
The Figure~\ref{ch1.sec4.subsec1.fig2}(b-2) confirms that over sufficiently large time, when $\lambda>0$, then the infectious population becomes extinct. Furthermore, note that the basic reproduction number in (\ref{ch1.sec2.lemma2a.corrolary1.eq4}) in this scenario is $R^{*}_{0}= 2.014926e-05<1$, which signifies that the disease is getting eradicated from the population over time, and the susceptible population seen in Figure~\ref{ch1.sec4.subsec1.fig2}(a-2) oscillates over sufficiently long time, and approaches the disease-free equilibrium state $S^{*}_{0}=\frac{B}{\mu}=0.03731667$.

Indeed, note that the minimum value for the sample path of $S(t)$ in Figure~\ref{ch1.sec4.subsec1.fig3}(a-2) over long time is $\min(S(t))=1.80286$. This, suggests that the susceptible population is asymptotically stable on average over sufficiently long time near $S^{*}_{0}=\frac{B}{\mu}=0.03731667$ as shown in Theorem~\ref{ch1.sec2b.thm2}. Also, note that the general decrease in the susceptible population $S(t)$ in Figure~\ref{ch1.sec4.subsec1.fig2}(a-2) over time is accounted for  by the presence of noise in the disease transmission rate with a significant intensity $\sigma_{\beta}=10$.

The Figure~\ref{ch1.sec4.subsec1.fig3} also obtained using the parameters of Table~\ref{ch1.sec4.table2},  and the initial conditions for the system (\ref{ch1.sec0.eq8})-(\ref{ch1.sec0.eq11}) in (\ref{ch1.sec4.eq1}), and with larger white noise intensity conditions for $\sigma_{\beta}=20$, and $\sigma_{i}=0,\forall i\in\{S, E, I, R\}$, confirms Theorem~\ref{ch1.sec2b.thm1} that the extinction of the infectious population $I(t)$ over sufficiently large time has no bearing on the size of the intensity $\sigma_{\beta}$, or the strength of the noise from the disease transmission rate, provided that the basic reproduction number $R^{*}_{0}<1$. For Figure~\ref{ch1.sec4.subsec1.fig3}, the basic reproduction number remains the same value $R^{*}_{0}= 2.014926e-05<1$ obtained for the scenario in Figure~\ref{ch1.sec4.subsec1.fig2}. Moreover, the estimate of the rate of extinction is also the same value of  $\lambda= 1.162485>0$ obtained for Figure~\ref{ch1.sec4.subsec1.fig2}.

In addition, the susceptible population in Figure~\ref{ch1.sec4.subsec1.fig3}(a-3) approaches the disease-free equilibrium state $S^{*}_{0}=\frac{B}{\mu}=0.03731667$, similarly to Figure~\ref{ch1.sec4.subsec1.fig2}(a-2), and the minimum value for the sample path of $S(t)$ in Figure~\ref{ch1.sec4.subsec1.fig3}(a-3) over long time is $\min(S(t))=1.918471$.

Therefore, from the behavior of the sample path for $S(t)$ in  Figure~\ref{ch1.sec4.subsec1.fig2}(a-2) and Figure~\ref{ch1.sec4.subsec1.fig3}(a-3), there is numerically evidence that the susceptible steady state population, $S^{*}_{0}=\frac{B}{\mu}$, is stochastically asymptotically stable, whenever the basic reproduction number $R^{*}_{0}<1$.
\section{Conclusion}
A stochastic family of SEIRS models for malaria is derived and studied, where the noises in the system represent the variability in the disease dynamics from the disease transmission and natural death rates. The threshold conditions for the extinction of malaria in the population over sufficiently long time are presented for both cases of (1) noises in the system from both the disease transmission and natural death rates, and (2) noise exclusively from the disease transmission rate.

The analytic results show that the dynamics of the disease in the case of noise exclusively from the disease transmission rate exhibits more profound characteristics such as (a) stability in the mean of the disease free steady state population asymptotically, (b) the threshold conditions for the extinction of malaria, and consequently for the asymptotic stability in the mean of the malaria-free population, are robust to the intensity of the noise from the disease transmission rate. Finally, numerical simulation results are presented to justify the analytical results of the study.
\newpage


\begin{thebibliography}{300}
\bibitem{macdonald} G. Macdonald, The analysis of infection rates in diseases in which superinfection occurs. Trop. Dis. Bull. 47, 907-915
\end{thebibliography}
\end{document}